\documentclass[%
 reprint,
 amsmath,amssymb,
 aps,
prx,
]{revtex4-2}

\usepackage{graphicx}
\usepackage{dcolumn}
\usepackage{bm}

\usepackage{amsmath}
\usepackage{amssymb}
\usepackage{booktabs}
\usepackage{hyperref}
\hypersetup{
    colorlinks,
    citecolor=blue,
    linkcolor=blue,
    urlcolor=blue
}

\usepackage[whole]{bxcjkjatype}
\usepackage[normalem]{ulem}
\usepackage{color}
\usepackage{soul}
\definecolor{americanrose}{rgb}{1.0, 0.01, 0.24}
\definecolor{electricpurple}{rgb}{0.75, 0.0, 1.0}
\definecolor{vividgreen}{rgb}{0.0, 0.6, 0.0}

\newcommand{\Eref}[1]{Eq.~\eqref{#1}}
\newcommand{\EEref}[1]{Equation~\eqref{#1}}
\newcommand{\Fref}[1]{Fig.~\ref{#1}}
\newcommand{\FFref}[1]{Figure~\ref{#1}}

\begin{document}

\preprint{APS/123-QED}

\title{Quantum Speed Limit and Quantum Thermodynamic Uncertainty Relation \\under Feedback Control}

\author{Hayato Yunoki}
 \email{yunoki@biom.t.u-tokyo.ac.jp}

\author{Yoshihiko Hasegawa}%
 \email{hasegawa@biom.t.u-tokyo.ac.jp}

\affiliation{
 Department of Information and Communication Engineering,\\
 Graduate School of Information Science and Technology,\\
 The University of Tokyo, Tokyo 113-8656, Japan
}


\begin{abstract}
Fundamental trade-off relations, such as quantum speed limit and quantum thermodynamic uncertainty relation, describe the performance limits of quantum systems. These relations impose that improving speed or precision of the quantum systems necessitates a substantial thermodynamic cost, thereby defining the ultimate operational limits. In practice, quantum feedback control, which is a pivotal technique for manipulating quantum dynamics based on measurement outcomes, is widely employed to enhance the performance of quantum systems. Nevertheless, how such active control protocols affect these fundamental performance bounds remains an open question. To elucidate the influence of feedback on these intrinsic limits, this work establishes a theoretical framework for quantum speed limit and quantum thermodynamic uncertainty relation under Markovian feedback protocol, which is a paradigmatic quantum feedback control. We derive general inequalities that incorporate the effects of feedback control on both speed and precision. Through numerical simulations on a simple two-level system and quantum error correction, a key application of quantum feedback control, we not only validate our derived bounds but also demonstrate that feedback control can indeed improve both speed and precision beyond those achievable limits in uncontrolled systems. Next, to elucidate the mechanism behind these improvements and the qualitative difference from uncontrolled dynamics, we conduct a detailed analysis of the governing thermodynamic costs, which are the fundamental quantities that constrain speed and precision, within a simple model. In particular, our analysis reveals that feedback can improve the time scaling order of these costs. This modification of the dynamical scaling is the origin of the qualitative performance gain, signifying that the feedback-induced improvements in speed and precision are not merely quantitative but represent a fundamental shift in the system's performance. Consequently, our work offers a comprehensive understanding of how feedback control impacts the fundamental limits on the speed and precision of quantum systems, providing crucial insights for the design of high-performance quantum technologies.
\end{abstract}

\maketitle

\section{Introduction}

The recent rapid development of quantum devices has promoted a growing interest in clarifying the fundamental performance limits of quantum systems. Trade-off relations provide a rigorous framework for revealing these limits, which ultimately distinguish what is achievable from what remains beyond reach. Thus, a comprehensive understanding of such relations is not merely a matter of foundational scientific curiosity but is also indispensable for guiding future technological advancements. The most well-known trade-off relation is Heisenberg's uncertainty principle \cite{heisenbergUberAnschaulichenInhalt1927, kennardZurQuantenmechanikEinfacher1927, robertsonUncertaintyPrinciple1929}, which states that non-commutative physical quantities such as position and momentum cannot be determined simultaneously with arbitrary precision in quantum systems. Building upon this foundational concept, investigations have been extended to trade-off relations governing the dynamical properties of quantum systems. Consequently, trade-off relationships that constrain the speed of state evolution and the precision of quantum processes have been studied intensively in recent years.

In 1945, Mandelstam and Tamm derived a lower bound on the time $\tau$ required for the quantum state of an isolated system to reach an orthogonal state \cite{mandelstam1945},
\begin{equation}\label{mtbound}
\tau \geq \frac{\pi}{2\mathrm{Std} (H)},
\end{equation}
where $\mathrm{Std} (H)$ is the standard deviation of the Hamiltonian $H$ and Dirac's constant $\hbar$ is set to $\hbar = 1$. \EEref{mtbound} represents a trade-off between time and energy fluctuation, implying that faster time evolution of a quantum system requires larger energy fluctuation. This type of inequality, which sets a limit on the speed of a quantum system's time evolution, is known as quantum speed limit (QSL) and has been generalized in various ways, with notable extensions to open quantum systems \cite{Mandelstam1991, margolusMaximumSpeedDynamical1998, deffnerGeneralizedClausiusInequality2010, taddeiQuantumSpeedLimit2013a, delcampoQuantumSpeedLimits2013, deffnerEnergyTimeUncertainty2013, piresGeneralizedGeometricQuantum2016, oconnorActionQuantumSpeed2021} (see \cite{deffnerQuantumSpeedLimits2017} for a review). QSL has also been extended to classical non-equilibrium systems, where it is referred to as classical speed limit \cite{okuyamaQuantumSpeedLimit2018, shanahanQuantumSpeedLimits2018, itoStochasticThermodynamicInterpretation2018, itoStochasticTimeEvolution2020, nicholsonTimeInformationUncertainty2020}. With the extension of QSL to open quantum and classical systems, it has become apparent that any acceleration in a system's evolution must be paid for by a greater thermodynamic cost, such as entropy production or dynamical activity, which counts average number of jumps of the system \cite{garrahanSimpleBoundsFluctuations2017, diterlizziKineticUncertaintyRelation2019}. Consequently, the concept of speed limit is now increasingly framed as a fundamental trade-off between speed and the thermodynamic cost \cite{shiraishiSpeedLimitClassical2018a, funoSpeedLimitOpen2019, voUnifiedApproachClassical2020, falascoDissipationTimeUncertaintyRelation2020, yoshimuraThermodynamicUncertaintyRelation2021, vanvuThermodynamicUnificationOptimal2023}. The importance of QSL is not confined to fundamental theory, but also imposes critical constraints on practical applications in quantum technology, such as quantum computing and quantum sensing \cite{lloydUltimatePhysicalLimits2000, howardImplementingTwoqubitGates2023, herbQuantumSpeedLimit2024}.

\begin{figure*}[t]
\includegraphics[width=180mm]{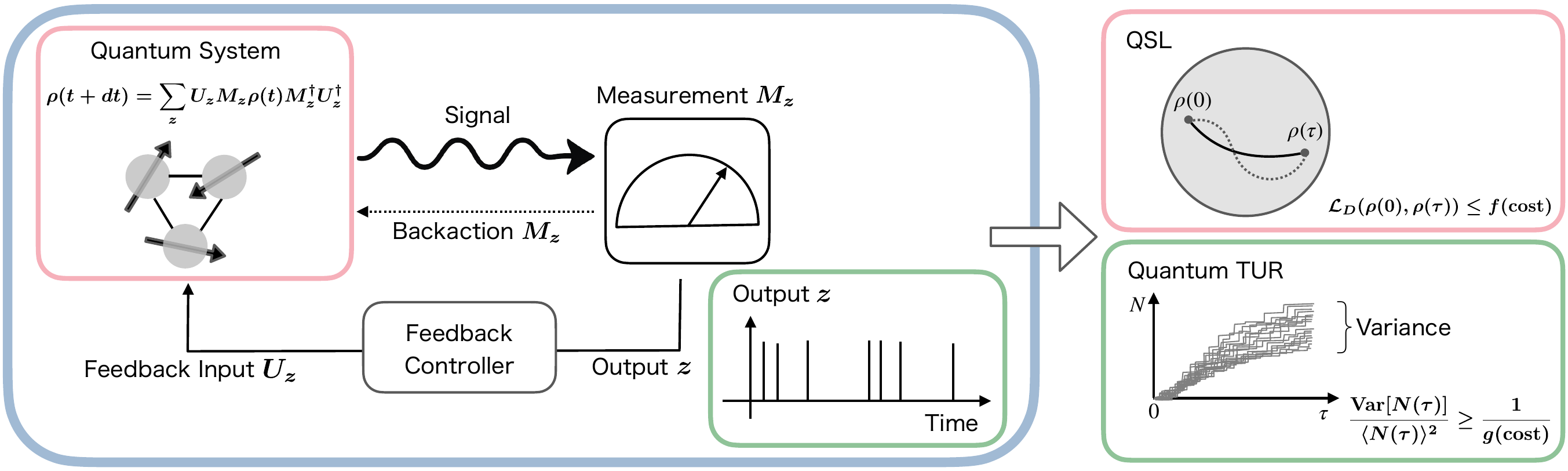}
\caption{\label{fig:fb}
Schematic illustration of our study. The left panel illustrates quantum feedback control loop, where quantum system, described by the density matrix $\rho(t)$ is measured by the Kraus operators $M_z$. Based on the measurement output $z$, a unitary feedback operation $U_z$ is applied to the system. Within this framework, we derive QSL and quantum TUR to clarify the contribution of feedback control to the speed and precision of the system. By focusing on the quantum system $\rho(t)$, we derive QSL, which represents that a lot of thermodynamic cost is required to increase the distance $\mathcal{L}_D$ that $\rho(t)$ travels during time interval $[0, \tau]$. By focusing on the output $z(t)$, we derive quantum TUR, which represents that a lot of thermodynamic cost is required to reduce the fluctuations of the measured quantity $N(\tau)\equiv \int_0^\tau dt z(t)$. 
} 
\end{figure*}

Interestingly, thermodynamics imposes fundamental limits not only on the speed of a process but also on its precision. Recently, the inequality that expresses this relation, called thermodynamic uncertainty relation (TUR), was discovered and has been actively explored in the field of classical stochastic thermodynamics \cite{baratoThermodynamicUncertaintyRelation2015, gingrichDissipationBoundsAll2016, garrahanSimpleBoundsFluctuations2017, dechantCurrentFluctuationsTransport2018, diterlizziKineticUncertaintyRelation2019, hasegawaUncertaintyRelationsStochastic2019, hasegawaFluctuationTheoremUncertainty2019, doi:10.1073/pnas.1918386117, koyukThermodynamicUncertaintyRelation2020, liuThermodynamicUncertaintyRelation2020} (see \cite{horowitzThermodynamicUncertaintyRelations2020} for a review). TUR sets a lower bound on the relative fluctuation (the noise-to-signal ratio) of an observable in terms of thermodynamic cost, thereby demonstrating that enhancing precision inevitably incurs a greater thermodynamic cost. For steady state, the TUR states
\begin{equation}
\frac{\mathrm{Var}[J]}{\langle J\rangle^2}\geq\frac{2}{\Sigma},
\end{equation}
where $\langle J\rangle$ and $\mathrm{Var}[J]$, respectively, are mean and variance of the current $J$ within the time interval $[0, \tau]$ and $\Sigma$ is entropy production \cite{baratoThermodynamicUncertaintyRelation2015}. TUR is applied to small systems that are strongly affected by fluctuations, for example, molecular motors and heat engines \cite{pietzonkaUniversalBoundEfficiency2016, pietzonkaUniversalTradeOffPower2018}. An alternative type of TUR gives a lower bound on the relative variance using dynamical activity $\mathcal{A}_c$ (cf. Eq.~\eqref{CDA_def}) instead of entropy production as follows, 
\begin{equation}
\frac{\mathrm{Var}[J]}{\langle J\rangle^2}\geq\frac{1}{\mathcal{A}_c}.
\end{equation}
This form of TUR based on dynamical activity is also known as kinetic uncertainty relation. TUR was originally established in classical systems, and has been extended to the quantum domain more recently, where it is known as quantum TUR \cite{erkerAutonomousQuantumClocks2017, brandnerThermodynamicBoundsPrecision2018,  carolloUnravelingLargeDeviation2019, liuThermodynamicUncertaintyRelation2019,  guarnieriThermodynamicsPrecisionQuantum2019, saryalThermodynamicUncertaintyRelation2019, hasegawaQuantumThermodynamicUncertainty2020, hasegawaThermodynamicUncertaintyRelation2021, sacchiThermodynamicUncertaintyRelations2021, kalaeeViolatingThermodynamicUncertainty2021,  hasegawaIrreversibilityLoschmidtEcho2021a, vanvuThermodynamicsPrecisionMarkovian2022, monnaiThermodynamicUncertaintyRelation2022, hasegawaUnifyingSpeedLimit2023, prechRoleQuantumCoherence2025, vanvuFundamentalBoundsPrecision2025}. It has been confirmed that classical TUR is violated in quantum regimes due to the effect of coherence, implying that coherence can enhance precision. The impact of quantum coherence on precision in TUR was actively studied in Ref. \cite{prechRoleQuantumCoherence2025}. The framework of quantum TUR has been actively investigated in the context of continuous measurement, which is a method of indirectly measuring a quantum system continuously over time (see \cite{landiCurrentFluctuationsOpen2024} for a review).

It is becoming clear that speed limit and TUR have a duality, and there are attempts to derive them in a unified manner, where dynamical activity plays a central role as the thermodynamic cost \cite{voUnifiedApproachClassical2020, hasegawaUnifyingSpeedLimit2023, hasegawaThermodynamicConcentrationInequalities2024}. For continuous measurement, QSL and quantum TUR can be derived in a unified manner by mapping the dynamics of the quantum system and the measurement results to the continuous matrix product state (cMPS), a technique known as the cMPS method \cite{hasegawaUnifyingSpeedLimit2023}. Within this formalism, focusing on the main system yields QSL, while focusing on the environment yields quantum TUR. The thermodynamic costs in these trade-off relations are characterized by quantum dynamical activity. Classical dynamical activity quantifies system's activity by based solely on jump statistics. Quantum dynamical activity quantifies the system's total activity by accounting not only for contributions from jumps but also for those from coherent dynamics. Classical dynamical activity is known to be closely related to classical Fisher information [\Eref{CDA}]. By analogy,  quantum dynamical activity was introduced as a quantity defined via quantum Fisher information [\Eref{QFI_DA}] in Ref.~\cite{hasegawaUnifyingSpeedLimit2023}. Subsequently, the exact solution for quantum dynamical activity in Lindblad dynamics was derived analytically in Refs.~\cite{nakajimaSymmetriclogarithmicderivativeFisherInformation2023, nishiyamaExactSolutionQuantum2024}.

Precise control is essential for maximizing the performance of physical systems. A cornerstone of modern engineering, feedback control provides a powerful mechanism for precisely guiding the behavior of complex and stable systems. This concept is adapted to the quantum realm as quantum feedback control: the manipulation of a quantum system using information obtained from its measurement results \cite{Wiseman_Milburn_2009, zhangQuantumFeedbackTheory2017} [\Fref{fig:fb}]. Due to the fragility of quantum systems to external disturbances, implementing feedback control is crucial for improving stability and realizing robust quantum technologies. Various experiments have been conducted with quantum feedback control \cite{dursoFeedbackCoolingOneElectron2003, gillettExperimentalFeedbackControl2010, zhouFieldLockedFock2012, inoueUnconditionalQuantumNoiseSuppression2013}. Given that standard projective measurements are maximally disruptive to a quantum system, quantum feedback control often employs continuous measurement that minimize back-action by extracting partial information over time \cite{wisemanQuantumTheoryOptical1993, wisemanQuantumTheoryContinuous1994, dohertyQuantumFeedbackControl2000, annby-anderssonQuantumFokkerPlanckMaster2022}. For example, quantum error correction is one of the main applications of quantum feedback control. Various methods have been proposed to detect and correct errors, corresponding to feedback control measurements and control inputs \cite{ahnContinuousQuantumError2002, ahnQuantumErrorCorrection2003, ahnQuantumErrorCorrection2004, sarovarPracticalSchemeError2004, borahMeasurementbasedEstimatorScheme2022}. Quantum feedback control has other applications such as preparation and stabilization of quantum states \cite{smithCaptureReleaseConditional2002, rossiMeasurementbasedQuantumControl2018, sayrinRealtimeQuantumFeedback2011, vijayStabilizingRabiOscillations2012}, metrology \cite{gefenParameterEstimationEfficient2016, zhouAchievingHeisenbergLimit2018, hongFeedbackControlQuantum2023}, and quantum batteries \cite{mitchisonChargingQuantumBattery2021, yaoStableChargingRydberg2021}. This technique also holds fundamental significance, particularly in quantum information theory and quantum thermodynamics. Maxwell's demon, which leverages measurement information to operate a system, serves as a paradigmatic example of feedback control \cite{leff2002maxwell}. Many works on Maxwell's demon have enabled an accurate formulation of the second law in a quantum feedback-controlled system \cite{sagawaSecondLawThermodynamics2008, morikuniQuantumJarzynskiSagawaUedaRelations2011, funoIntegralQuantumFluctuation2013, yadaQuantumFluctuationTheorem2022, prechQuantumFluctuationTheorem2024}. Thermodynamics in quantum feedback-controlled systems has also been actively discussed \cite{10.1143/PTP.127.1, gooldRoleQuantumInformation2016, prechQuantumThermodynamicsContinuous2025}.

Providing a powerful toolkit for steering the dynamics of quantum systems, quantum feedback control naturally raises the question of whether it can be harnessed to enhance their speed and precision. However, prior investigations into this topic have been limited in scope. For instance, while the framework of QSL is applicable to general Markovian open quantum systems, its extension to scenarios involving feedback control has thus far been confined to a two-level atom \cite{yuQuantumSpeedLimit2018}. In addition, quantum TUR under feedback control has been proposed in Ref.~\cite{hasegawaQuantumThermodynamicUncertainty2023}, which is a preliminary and unpublished preprint by one of the authors. In that work, the quantum dynamical activity which is the thermodynamic cost that bounds precision is defined by the abstract mathematical quantity that is not expressed solely in terms of the physical observables of the system of interest. Consequently, it remains unclear how the precision is affected by feedback control. Furthermore, the above duality between the QSL and quantum TUR has not yet been explored in the presence of feedback control.

\begin{table*}
\caption{\label{result_table}Summary of our results. QSL, quantum TUR derived by cMPS method, quantum TUR derived by concentration inequality, and quantum dynamical activity as a cost in these trade-off relations under feedback control by jump, homodyne, Gaussian measurement. 
}
\begin{ruledtabular}
{
\renewcommand{\arraystretch}{1.8}
\begin{tabular}{@{}lllll@{}} \toprule
        & Jump Measurement & Homodyne (Gaussian) Measurement \\ \midrule
\begin{tabular}[c]{@{}l@{}}QSL  \\ (cMPS method)  \end{tabular}                                                                                   &     $\displaystyle \mathcal{L}_D(\rho(0), \rho(\tau))\leq\frac{1}{2}\int_0^\tau dt \frac{\sqrt{\mathcal{B}_{\mathrm{jmp}}^{\mathrm{fb}}(t)}}{t}$             &       $\displaystyle \mathcal{L}_D(\rho(0), \rho(\tau))\leq\frac{1}{2}\int_0^\tau dt \frac{\sqrt{\mathcal{B}_{\mathrm{hom(gau)}}^{\mathrm{fb}}(t)}}{t}$               \\ \hline
\begin{tabular}[c]{@{}l@{}}Quantum TUR\\ (cMPS method)\end{tabular} &      $\displaystyle \frac{\mathrm{Var}[N(\tau)]}{\langle N(\tau) \rangle^2}\geq \frac{1}{\mathcal{B}_{\mathrm{jmp}}^{\mathrm{fb}}(\tau)}$            &       $\displaystyle \frac{\mathrm{Var}[Z(\tau)]}{\langle Z(\tau) \rangle^2}\geq \frac{1}{4\mathcal{B}_{\mathrm{hom(gau)}}^{\mathrm{fb}}(\tau)}$               \\ \hline
\begin{tabular}[c]{@{}l@{}}Quantum TUR\\ (Concentration Inequality)\end{tabular}               &    $\displaystyle \frac{\|N^\circ(\tau)\|_p}{\|N^\circ(\tau)\|_1}\geq \sin{\left[\frac{1}{2}\int_0^\tau dt\frac{\sqrt{\mathcal{B}_{\mathrm{jmp}}^{\mathrm{fb}}(t)}}{t}\right]}^{-\frac{2(p-1)}{p}}$                 &                      \\ \hline
Quantum Dynamical Activity                                                                     &   $ \displaystyle \mathcal{B}_{\mathrm{jmp}}^{\mathrm{fb}}(\tau)$               &         $\displaystyle \mathcal{B}_{\mathrm{hom(gau)}}^{\mathrm{fb}}(\tau)$            \\ \hline

\end{tabular}
}
\end{ruledtabular}
\end{table*}

In this paper, we derive QSL and quantum TUR under Markovian feedback control based on continuous measurement [\Fref{fig:fb}]. Our framework is applicable to any open quantum system undergoing Markovian evolution, and the quantum dynamical activity that constrains both speed and precision is derived as an exact expression formulated solely in terms of the physical quantities of the system itself. Furthermore, using these results, we provide a detailed discussion, absent in existing studies, on how feedback control impacts both speed and precision.

We consider three forms of continuous measurements: jump measurement, homodyne measurement (also known as diffusion measurement), and Gaussian measurement. These measurement schemes are typical for Markovian feedback control, as introduced in Refs.~\cite{wisemanQuantumTheoryOptical1993, wisemanQuantumTheoryContinuous1994, annby-doc}. In this paper, QSL and quantum TUR under these cases are derived based on the cMPS method, and quantum dynamical activity under these feedback control, which represent the thermodynamic costs, are derived analytically based on Refs. \cite{nakajimaSymmetriclogarithmicderivativeFisherInformation2023, nishiyamaExactSolutionQuantum2024}. In particular, for the quantum dynamical activity under jump measurement based feedback control, we performed a detailed analysis using Taylor expansion to isolate and evaluate the contribution originating from the feedback control. In addition, we also focus on another type of quantum TUR by using concentration inequality derived in Ref.~\cite{hasegawaThermodynamicConcentrationInequalities2024}. This bound is based on continuous measurement by jump measurement and the same quantum dynamical activity as a thermodynamic cost. We also extend this type of quantum TUR to incorporate feedback control by jump measurement (see Table \ref{result_table} for summary of our results).

By adapting our results to two-level atom driven by a classical laser field and performing numerical simulations, derived inequalities are verified and the contribution of feedback control to speed and precision is investigated. These results show that feedback control can increase the speed and improve the precision of the quantum system. In addition, we focus on an important application of feedback control: quantum error correction, which is an essential tool for quantum computation. In particular, we adapt our result to continuous quantum error correction, in which quantum error correction is performed by feedback input based on the results of continuous measurements \cite{mabuchiInversionQuantumJumps1996, ahnContinuousQuantumError2002, ahnQuantumErrorCorrection2003, gefenParameterEstimationEfficient2016, atalayaContinuousQuantumError2021, livingstonExperimentalDemonstrationContinuous2022}. In typical quantum error correction, a projective measurement is made at each discrete time round, and a unitary gate corresponding to the error syndrome is applied. On the other hand, continuous quantum error correction constantly monitors errors and provides real-time Hamiltonian feedback input when the error is detected. Continuous quantum error correction is useful when using time-dependent Hamiltonian such as quantum annealing and quantum simulation. It also allows for the reduction of ancillary qubits and entangled gates compared to typical quantum error correction. Also in quantum error correction, we numerically confirm the validity of the derived quantum TUR and demonstrate that feedback enhances precision. Finally, to reveal the underlying mechanism of the feedback-induced improvements of speed and precision, we conduct a detailed analysis of the quantum dynamical activity under feedback control, the central thermodynamic cost governing the QSL and quantum TUR in our framework. The behavior of this quantity dictates the fundamental limits on these performance metrics for quantum systems under feedback control. We elucidate the properties of the quantity, clarifying the differences arising from the presence or absence of feedback, as well as from the choice of measurement scheme. To illustrate the properties, we study a simple two-level atom driven by a classical laser field and calculate the quantum dynamical activity as a function of time and feedback strength. In particular, this analysis reveals a key finding that feedback control can qualitatively alter the time scaling order of quantum dynamical activity, a fundamental departure from the behavior in uncontrolled systems. Our results demonstrate that the feedback-induced improvements in speed and precision are not merely quantitative adjustments but originate from a fundamental reshaping of the system's thermodynamic landscape. Ultimately, this work provides a rigorous and unified understanding of how information gain and control actions shape the fundamental performance limits of quantum systems.

\section{Methods}
\subsection{QSL and Quantum TUR without Feedback Control}
We begin by reviewing QSL and quantum TUR without feedback control, that is, a system subject only to continuous measurement \cite{hasegawaQuantumThermodynamicUncertainty2020, hasegawaUnifyingSpeedLimit2023}. The Markovian dynamics of an open quantum system is governed by the Lindblad (GKSL) equation \cite{goriniCompletelyPositiveDynamical1976a, lindbladGeneratorsQuantumDynamical1976}: 
\begin{equation}\label{Lindblad}
\begin{split}
    \frac{d\rho(t)}{dt}=&\mathcal{L}\rho(t)\\
    =&\mathcal{H}\rho(t)+ \sum_{z=1}^{N_c}\mathcal{D}[L_z]\rho(t),
\end{split}
\end{equation}
where $\rho(t)$ is the density operator of the system at time $t$, $\mathcal{L}$ is the Lindblad superoperator, and $\mathcal{H}\rho\equiv-i[H, \rho]$ describes unitary time evolution with the system Hamiltonian $H$. $\mathcal{D}[L]\rho=L\rho L^\dag-\frac{1}{2}\{L^\dag L, \rho\}$ is the dissipator, where $\{A, B\}\equiv AB+BA$ is the anti-commutator, $L_z$ is the jump operator, and $N_c$ is the number of channels (or jump operator). 

Lindblad equation is also known to describe the continuously measured system \cite{landiCurrentFluctuationsOpen2024}. We now elucidate this measurement-based picture by unraveling the equation into stochastic quantum trajectories. When the dynamics is governed by \Eref{Lindblad}, the time evolution of $\rho(t)$ over an infinitesimal time $dt$ can be described by
\begin{equation}\label{Lindblad_Kraus}
\rho(t+dt)=e^{\mathcal{L}dt}\rho(t)=\sum_{z=0}^{N_c}M_z\rho_c(t)M_z^\dag,
\end{equation}
where $M_0=\mathbb{I}-iH_{\mathrm{eff}}dt$ and $M_z=\sqrt{dt}L_z$.  Here, $\mathbb{I}$ denotes identity operator, and $H_{\mathrm{eff}}\equiv H-\frac{i}{2}\sum_{z=1}^{N_c}L_z^\dag L_z$ is non-Hermitian effective Hamiltonian. \EEref{Lindblad_Kraus} represents the Kraus representation of the measurement process, where the corresponding Kraus operator is $M_z$. $M_z (z=1, 2, ..., N_c)$ indicates that a discontinuous jump corresponding to $L_z$ has occurred, whereas $M_0$ signifies no jump. A jump corresponding to $M_z (z=1, 2, ..., N_c)$ is detected with probability $p_z(t)=\mathrm{Tr}[M_z\rho_c(t)M_z^\dag]$ during the infinitesimal time interval $dt$, where $\rho_c(t)$ is the density operator conditioned on the results of the previous measurement at time $t$. The conditional time evolution when a jump corresponding to $M_z$ is detected is given by
\begin{equation}\label{unraveling_lindblad}
\rho_c(t+dt)=\frac{M_z\rho_c(t)M_z^\dag}{p_z(t)}.
\end{equation}
\EEref{unraveling_lindblad} is known as the unraveling of Lindblad equation, which maps the master equation onto a stochastic trajectory of the density operator such that averaging over these trajectories recovers the original master equation. By considering the detection of jumps corresponding to $M_z$ as continuous in time, we can consider continuous measurement by jump measurement. When performing continuous measurement by jump measurement, we introduce Poisson increment $dN_z$. $dN_z$ takes the value 1 when a jump corresponding to $M_z$ is detected in $dt$, and 0 otherwise. Using $dN_z$, the output current $I(t)$ is defined as
\begin{equation}\label{jump_current}
    I(t)\equiv\sum_z \nu_z\frac{dN_z}{dt},
\end{equation}
where $\nu_z$ is the weight associated with each jump.

Because the Kraus representation is not unique, Lindblad equation \Eref{Lindblad} remains invariant under the following transformation:
\begin{equation}\label{homodyne_transform}
    L_z\rightarrow L_z+\alpha_z, H\rightarrow H-\frac{i}{2}\sum_z(\alpha_z^*L_z-\alpha_z L_z^\dagger),
\end{equation}
where $\alpha_z=|\alpha_z|e^{i\phi_z}$ are arbitrary complex constants. By taking the limit where $|\alpha_z|$ becomes large, we can consider continuous measurement by homodyne measurement.  In homodyne measurements, the output current $I_{\mathrm{hom}}(t)$ is defined by
\begin{equation}\label{homodyne_current}
    I_{\mathrm{hom}}(t)\equiv \sum_z\nu_z\left(\langle e^{-i\phi_z}L_z+e^{i\phi_z}L_z^\dagger\rangle+\frac{dW_z}{dt}\right ),
\end{equation}
where $\langle\bullet\rangle$ is the expectation of operator $\bullet$ and $dW_z$ is Wiener increment. Wiener increment follows Gaussian distribution with $\langle dW_z\rangle=0$ and $\langle dW_z^2\rangle=dt$. We can consider continuous measurement by Gaussian measurement, where the measurement operator is given by
\begin{equation}\label{gaussian_kraus}
    M_z=\left(\frac{2\lambda dt}{\pi}\right)^{1/4}e^{-\lambda dt (I_{\mathrm{gau}}-Y)^2},
\end{equation}
where $I_{\mathrm{gau}}\in \mathbb{R}$ denotes the measurement output, $\lambda>0$ is the measurement strength, and $Y$ is an Hermitian observable \cite{jacobsStraightforwardIntroductionContinuous2006, landiCurrentFluctuationsOpen2024, annby-doc}. From a physical perspective, Gaussian measurement corresponds to the measurement of an observable $Y$ subjected to Gaussian noise, which realistically models the random fluctuations inherent in a measurement apparatus. This process involves a fundamental trade-off governed by the measurement strength $\lambda$. Specifically, a smaller $\lambda$ minimizes the back-action on the system, thereby preserving its quantum coherence, but at the expense of increased fluctuations in the measured current. In the strong measurement limit, where $\lambda dt \rightarrow \infty$, the Gaussian measurement reduces to a projective measurement. The probability of obtaining the measurement output $I_{\mathrm{gau}}$ is given by
\begin{equation}
\text{Tr} \left\{ M_z \rho M_z^\dagger \right\} = \left( \frac{2\lambda dt}{\pi} \right)^{1/2} \sum_y e^{-2\lambda dt (I_{\mathrm{gau}}-y)^2} \langle y | \rho | y \rangle,
\end{equation}
where $y$ and $|y\rangle$ are the eigenvalue and eigenvector of $Y$, respectively. This is a Gaussian distribution with a mean of $y$ and a standard deviation of $1/(2\sqrt{\lambda dt})$. The output current of each measurement is described by 
\begin{equation} \label{gaussian_current}
    I_{\mathrm{gau}}(t)=\langle Y\rangle + \frac{1}{2\sqrt{\lambda}}\frac{dW}{dt},
\end{equation}
where $dW$ is Wiener increment. The unraveling and output current of Gaussian measurement coincide with those of homodyne measurement by setting jump operator $L=\sqrt{\lambda}Y$, $\phi_z=0$, and weight $\nu=1/(2\sqrt{\lambda})$.

To derive QSL and quantum TUR, mapping the Lindblad dynamics to cMPS is a common approach \cite{hasegawaQuantumThermodynamicUncertainty2020, hasegawaUnifyingSpeedLimit2023, hasegawaIrreversibilityLoschmidtEcho2021a, hasegawaThermodynamicUncertaintyRelation2022}. MPS has been applied to explore Markov processes in stochastic and quantum thermodynamics \cite{garrahanThermodynamicsQuantumJump2010, lesanovskyCharacterizationDynamicalPhase2013, garrahanClassicalStochasticDynamics2016}.  When continuous measurements are performed over the time interval [0, $\tau$] and this interval is discretized into large $N_\tau$ subdivisions, the states of the system and the environment are described by MPS :
\begin{equation}\label{MPS}
    |\Psi(\tau)\rangle=\sum_{z_0, ..., z_{N_\tau-1}}M_{z_{N_\tau-1}}\cdots M_{z_0}|\psi(0)\rangle\otimes|z_{N_\tau-1}, \cdots z_{0}\rangle.
\end{equation}
This pure state $|\Psi(\tau)\rangle$ has all the information of the continuous measurement process. The environment state $|z_{N_\tau-1}, \cdots z_{0}\rangle$ represents the measurement outcomes. In the limit of sufficiently large $N_\tau$, the state $|\Psi(\tau)\rangle$ converges to cMPS \cite{verstraeteContinuousMatrixProduct2010, osborneHolographicQuantumStates2010}.

Quantum Fisher information plays a central role in the derivation of both QSL and quantum TUR for continuously measured systems. Quantum Fisher information $\mathcal{J}(\xi)$ for a pure state $|\psi(\xi)\rangle$ parametrized by $\xi$ is given by \cite{meyerFisherInformationNoisy2021}
\begin{equation}
\mathcal{J}(\xi)=4\left[\langle\partial_\xi\psi(\xi)|\partial_\xi\psi(\xi)\rangle-|\langle\partial_\xi\psi(\xi)|\psi(\xi)\rangle|^2\right].
\end{equation}
By parametrizing $L_z$ and $H$ as $L_z(\xi)$ and $H(\xi)$, the cMPS $|\Psi(\tau)\rangle$ can be extended to a parametrized form $|\Psi(\tau, \xi)\rangle$ for Lindblad dynamics. This formulation enables the quantum Fisher information for continuous measurements to be expressed as
\begin{equation}\label{cMPS_QFI}
    \mathcal{I}(\xi)= 4\left[\langle\partial_\xi\Psi(\tau, \xi)|\partial_\xi\Psi(\tau, \xi)\rangle-|\langle\partial_\xi\Psi(\tau, \xi)|\Psi(\tau, \xi)\rangle|^2\right].
\end{equation}
Here, the quantity $\mathcal{I}(\xi)$ have the information of the parametrized Lindblad dynamics over the interval $[0,\tau]$. As we will subsequently demonstrate, it provides the basis for defining the thermodynamic cost that appears in the trade-off relations we aim to establish.

When considering the duality between speed limit and TUR, dynamical activity emerges as the key thermodynamic cost. Considering a classical Markov process, classical dynamical activity $\mathcal{A}_{\mathrm{c}}(\tau)$ is defined as \cite{maesFrenesyTimesymmetricDynamical2020}
\begin{equation}\label{CDA_def}
    \mathcal{A}_{\mathrm{c}}(\tau)\equiv \int_0^\tau dt \sum_{\nu, \mu (\nu\neq\mu)}P_\mu(t)W_{\nu\mu},
\end{equation}
where $P_\mu(t)$ denotes the probability of being the state $\mu$ at time $t$, and $W_{\nu\mu}$ represents the transition rate from state $\mu$ to $\nu$. Classical dynamical activity is a thermodynamic quantity that quantifies the average number of jumps within the time interval [0, $\tau$]. When the Markov process is parameterized by a time-scaling parameter $\xi=t/\tau$, the classical Fisher information $\mathcal{I}_\mathrm{c}(\xi)$ associated with the parametrized state satisfies
\begin{equation}\label{CDA}
    \mathcal{I}_\mathrm{c}(\xi)=\frac{\mathcal{A}_{\mathrm{c}}(t)}{\xi^2}.
\end{equation}
In Ref.~\cite{hasegawaUnifyingSpeedLimit2023}, quantum dynamical activity $\mathcal{B}(t)$ is defined by
\begin{equation}\label{QFI_DA}
    \mathcal{B}(t)\equiv \xi^2\mathcal{I}(\xi)
\end{equation}
as the quantum counter part of \Eref{CDA}, where cMPS $|\Psi(\tau, \xi)\rangle$ is scaled by $t/\tau$ with respect to time. In the case of $\xi=1$, $\mathcal{B}(\tau)$ is obtained.

In the formulation of QSL, we adopt Bures distance $\mathcal{L}_D$ to quantify the distance between quantum states, defined as follows:
\begin{equation}
    \mathcal{L}_D(\rho_1, \rho_2)\equiv\mathrm{arccos}\sqrt{\mathrm{Fid}(\rho_1, \rho_2)}, 
\end{equation}
with quantum fidelity $\mathrm{Fid}(\rho_1, \rho_2)$ given by \cite{Nielsen_Chuang_2010}
\begin{equation}
    \mathrm{Fid}(\rho_1, \rho_2)\equiv \left(\mathrm{Tr}\sqrt{\sqrt{\rho_1}\rho_2\sqrt{\rho_2}}\right)^2.
\end{equation}
When considering the distance between two pure states $|\psi(t_1)\rangle$ and $|\psi(t_2)\rangle$, its upper bound is given by \cite{taddeiQuantumSpeedLimit2013a}
\begin{equation}\label{geometric_QSL}
    \mathcal{L}_D(|\psi(t_1)\rangle, |\psi(t_2)\rangle)\leq \frac{1}{2}\int_{t_1}^{t_2} dt\sqrt{\mathcal{J}(t)}.
\end{equation}
\EEref{geometric_QSL} is known as geometric quantum speed limit. In Ref. \cite{hasegawaUnifyingSpeedLimit2023}, QSL for Lindblad dynamics (continuous measurements) is derived by using the geometric QSL [\Eref{geometric_QSL}] as follows. When considering geometric QSL for cMPS, we obtain
\begin{equation}\label{cMPS_QSL}
    \mathcal{L}_D(|\Psi(0)\rangle, |\Psi(\tau)\rangle)\leq\frac{1}{2}\int_0^\tau dt \sqrt{\mathcal{I}(t)}.
\end{equation}
Bures distance satisfies monotonicity property
\begin{equation}
    \mathcal{L}_D(\rho_1, \rho_2)\geq\mathcal{L}_D(\varepsilon(\rho_1), \varepsilon(\rho_2))
\end{equation}
for any completely positive and trace-preserving (CPTP) map $\varepsilon$. For \Eref{cMPS_QSL}, tracing out the environment for cMPS [\Eref{MPS}] and using the relationship in \Eref{QFI_DA}, we obtain
\begin{equation}\label{QSL}
    \mathcal{L}_D(\rho(0), \rho(\tau))\leq\frac{1}{2}\int_0^\tau dt \frac{\sqrt{\mathcal{B}(t)}}{t},
\end{equation}
since tracing out the environment corresponds to CPTP map. \EEref{QSL} thus represents QSL for Lindblad dynamics.

By parametrizing Lindblad dynamics as
\begin{equation}\label{lindblad_scaling}
    H(\theta)=(1+\theta)H, L_z(\theta)=\sqrt{1+\theta}L_z,
\end{equation}
we can introduce a time scaling factor of $\xi=1+\theta$ for the corresponding cMPS. Reference \cite{hasegawaQuantumThermodynamicUncertainty2020} derived quantum TUR for continuous measurement by applying quantum Cram\'er-Rao inequality to the quantum Fisher information obtained from the cMPS $\mathcal{I}(\theta)$. Quantum Cram\'er-Rao inequality states \cite{hottaQuantumEstimationLocal2004}
\begin{equation}\label{cramer}
    \frac{\mathrm{Var}_\theta[\Theta]}{(\partial_\theta\langle\Theta\rangle_\theta)^2}\geq\frac{1}{\mathcal{I}(\theta)},
\end{equation}
where $\Theta$ is an observable with the measurement. To establish quantum TUR for dynamics over the time interval [0, $\tau$], we define the observable for jump measurement based on the current $I(t)$ [\Eref{jump_current}] as follows:
\begin{equation}\label{Ntau_def}
    N(\tau)\equiv\int_0^\tau dt I(t) = \sum_z \nu_zN_z(\tau)
\end{equation}
where $N_z(\tau)$ denotes the number of the jumps associated with the measurement operator $M_z$ occurring within [0, $\tau$]. In this parametrization [\Eref{lindblad_scaling}], quantum dynamical activity $\mathcal{B}(\tau)$ is given by 
\begin{equation}\label{lindblad_QDA}
    \mathcal{B}(\tau)=\mathcal{I}(\theta)|_{\theta=0}.
\end{equation}
The dynamics of $\theta=0$ reduces to the original dynamics, we can obtain
\begin{equation}
    \mathrm{Var}_{\theta=0}[N(\tau)]= \mathrm{Var}[N(\tau)].
\end{equation}
By setting $1+\theta=t/\tau$, the condition $\theta=0$ corresponds to $t=\tau$, which yields
\begin{equation}\label{exp_scaling}
    \partial_\theta\langle N(\tau)\rangle_\theta|_{\theta=0}=\tau\partial_t\left\langle N(\tau)\right\rangle_t|_{t=\tau}=\tau\partial_\tau\langle N(\tau)\rangle.
\end{equation}
From \Eref{cramer}-(\ref{exp_scaling}), the following quantum TUR is obtained:
\begin{equation}\label{lindblad_jump_tur}
    \frac{\mathrm{Var}[N(\tau)]}{\tau^2(\partial_\tau\langle N(\tau) \rangle)^2}\geq \frac{1}{\mathcal{B}(\tau)}.
\end{equation}
For the steady state condition, \Eref{exp_scaling} becomes
\begin{equation}
    \partial_\theta\langle N(\tau)\rangle_\theta|_{\theta=0}=\partial_\theta (1+\theta)\left\langle N(\tau)\right\rangle_{\theta=0}|_{\theta=0}=\langle N(\tau)\rangle.
\end{equation}
Consequently, for the steady state, the following quantum TUR holds:
\begin{equation}\label{lindblad_jump_tur_steady}
    \frac{\mathrm{Var}[N(\tau)]}{\langle N(\tau) \rangle^2}\geq \frac{1}{\mathcal{B}(\tau)}.
\end{equation}
When performing conituous measurement by homodyne measurement, we define the observable $Z(\tau)$ for dynamics over [0, $\tau$] as
\begin{equation}\label{hom_obs}
    Z(\tau)\equiv\int_0^\tau dt I_{\mathrm{hom}}(t),
\end{equation}
where $I_\mathrm{hom}(t)$ is the measurement current for homodyne measurement, as defined in \Eref{homodyne_current}. In this case, the following relation holds:
\begin{equation}
    \partial_\theta\langle Z(\tau)\rangle_\theta|_{\theta=0}=\partial_\theta \sqrt{1+\theta}\left\langle Z(\tau)\right\rangle_{\theta=0}|_{\theta=0}=\frac{\langle Z(\tau)\rangle}{2}.
\end{equation}
Thus, quantum TUR for homodyne measurement under  steady state condition takes the form
\begin{equation}\label{hom_tur}
    \frac{\mathrm{Var}[Z(\tau)]}{\langle Z(\tau) \rangle^2}\geq \frac{1}{4\mathcal{B}(\tau)}.
\end{equation}
Leveraging the correspondence between Gaussian measurement and homodyne measurement, we can define the observable $Z(\tau)$ with $I_\mathrm{gau}(t)$ in a manner similar to \Eref{hom_obs}. Consequently, the quantum TUR for Gaussian measurement under steady state condition become the same form as that in \Eref{hom_tur}.

Additionally, Ref. \cite{hasegawaThermodynamicConcentrationInequalities2024} derived alternative formulation of the quantum TUR based on concentration inequalities \cite{boucheronConcentrationInequalitiesNonasymptotic2013, CMR-37-1} rather than employing cMPS and Cram\'er-Rao inequality, particularly in the context of jump measurements.
For $p>1$ and $0\leq(1/2)\int_0^\tau dt \sqrt{\mathcal{B}(t)}/t\leq\pi/2$, the following bound holds:
\begin{equation}\label{concentration_QTUR}
    \frac{\|N^{\circ}(\tau)\|_p}{\|N^{\circ}(\tau)\|_1}\geq \sin{\left[\frac{1}{2}\int_0^\tau dt\frac{\sqrt{\mathcal{B}(t)}}{t}\right]}^{-\frac{2(p-1)}{p}},
\end{equation}
where $\|\cdot\|_p$ denotes the $p$-norm, defined as $\|\cdot\|_p\equiv\langle |\cdot|^p \rangle^{1/p}$. Please note that the observable $N^\circ(\tau)$ in Eq.~\eqref{concentration_QTUR} is defined more generally than $N(\tau)$ in Eq.~\eqref{Ntau_def}. In particular, $N^\circ(\tau)$ can be any function of the jump records, provided that $N^\circ(\tau) = 0$ when no jumps occur. For $p=2$, the left-hand side can be expressed in the form of variance over the square of the mean, allowing quantum TUR to be recovered. \EEref{concentration_QTUR} generalizes the quantum TUR previously obtained in Ref. \cite{hasegawaUnifyingSpeedLimit2023}.


\subsection{Exact Quantum Dynamical Activity without Feedback Control}
For Lindblad dynamics, time scaling by a factor of $(1+\theta)$ is performed by parametrizing of \Eref{lindblad_scaling}, and under this parametrization, quantum dynamical activity is defined by \Eref{lindblad_QDA}. Since quantum dynamical activity $\mathcal{B}(\tau)$ is currently defined only via quantum Fisher information $\mathcal{I}(\xi)$ [\Eref{cMPS_QFI}], a direct physical interpretation is challenging. For this reason, we consider obtaining a formulation of $\mathcal{B}(\tau)$ based solely on tangible physical quantities of the system. To derive an exact expression for \Eref{lindblad_QDA}, we should calculate $\mathcal{I}(\theta)|_{\theta=0}$. Focusing on \Eref{cMPS_QFI}, a key step is to evaluate the following:
\begin{equation}
\begin{split}
    |\langle\Psi(\tau, \phi)|\Psi(\tau, \theta)\rangle|=&\mathrm{Tr}_{SE}[|\Psi(\tau, \theta)\rangle\langle\Psi(\tau, \phi)|]\\
    =&\mathrm{Tr}_S\left[\mathrm{Tr}_E[|\Psi(\tau, \theta)\rangle\langle\Psi(\tau, \phi)|]\right].
\end{split}
\end{equation}
 When we define 
\begin{equation}\label{two_rho}
    \rho^{\theta, \phi}(\tau)\equiv\mathrm{Tr}_E[|\Psi(\tau, \theta)\rangle\langle\Psi(\tau, \phi)|],
\end{equation}
the following relation holds:
\begin{equation}
    \rho^{\theta, \phi}(t+dt)=\sum_z M_z(\theta)\rho^{\theta, \phi}(t)M_z^\dag(\phi).
\end{equation}
Noting that $\rho^{\theta,\phi}$ is not density operator. Thus, $\rho^{\theta, \phi}$ satisfies two-sided Lindblad equation \cite{gammelmarkFisherInformationQuantum2014a}:
\begin{equation}
\begin{split}
     \frac{d\rho^{\theta, \phi}}{dt}=&\mathcal{L}^{\theta, \phi}\rho^{\theta, \phi}\\
    =&\mathcal{H}(\theta, \phi)\rho^{\theta, \phi}+ \sum_{z=1}^{N_c} \Big[L_z(\theta)\rho^{\theta, \phi}L_z^\dag(\phi)\\
    &-\frac{1}{2}\{L_z^\dag(\theta) L_z(\theta)\rho^{\theta, \phi}+\rho^{\theta, \phi}L_z^\dag(\phi) L_z(\phi)\}\Big],
\end{split}
\end{equation}
where $\mathcal{H}(\theta, \phi)\rho^{\theta, \phi}\equiv-i[H(\theta)\rho^{\theta, \phi}-\rho^{\theta, \phi}H(\phi)]$. Quantum dynamical activity can also be expressed as
\begin{equation}
    \mathcal{B}(\tau)=4[\partial_\theta \partial_\phi C(\theta, \phi)-\partial_\theta C(\theta, \phi)\partial_\phi C(\theta, \phi)]|_{\theta=\phi=0},
\end{equation}
where $C(\theta, \phi)\equiv \mathrm{Tr}_S \rho^{\theta, \phi}(\tau)$. Reference \cite{hasegawaUnifyingSpeedLimit2023, hasegawaQuantumThermodynamicUncertainty2020} defined quantum dynamical activity in this manner, but its analytical solution had not been clarified. Reference \cite{nakajimaSymmetriclogarithmicderivativeFisherInformation2023, nishiyamaExactSolutionQuantum2024} recently provided an exact analytical representation of quantum dynamical activity for Lindblad dynamics.

Nakajima and Utsumi derived the following expression \cite{nakajimaSymmetriclogarithmicderivativeFisherInformation2023}:
\begin{equation}\label{Nakajima_QDA}
    \mathcal{B}(\tau) =\mathcal{A}(\tau) + 4(I_1 + I_2) - 4 \left( \int_0^\tau ds \, \mathrm{Tr}_S[H\rho(s)] \right)^2, 
\end{equation}
where 
\begin{equation}
\mathcal{A}(\tau)\equiv \int_0^\tau dt \sum_z \mathrm{Tr}_S[L_z\rho(t)L_z^\dag],
\end{equation}
\begin{equation}
I_1 \equiv \int_0^\tau ds_1 \int_0^{s_1} ds_2 \, \mathrm{Tr}_S\left[\mathcal{K}_2 e^{\mathcal{L}(s_1 - s_2)} \mathcal{K}_1 \rho(s_2)\right],
\end{equation}
\begin{equation}
I_2 \equiv \int_0^\tau ds_1 \int_0^{s_1} ds_2 \, \mathrm{Tr}_S\left[\mathcal{K}_1 e^{\mathcal{L}(s_1 - s_2)} \mathcal{K}_2 \rho(s_2)\right],
\end{equation}
with
\begin{equation}
    \mathcal{K}_1 \bullet \equiv -iH_\mathrm{eff} \bullet + \frac{1}{2} \sum_z L_z \bullet L_z^\dagger,
\end{equation}
\begin{equation}
    \mathcal{K}_2 \bullet \equiv i \bullet H_\mathrm{eff}^\dagger + \frac{1}{2} \sum_z L_z \bullet L_z^\dag.
\end{equation}
$\mathcal{A}(\tau)$ quantifies the number of jumps in the time interval [0, $\tau$], corresponding to a direct extension of the classical dynamical activity [\Eref{CDA_def}]. This follows from the fact that Lindblad equation describes classical Markov process when Hamiltonian $H=0$ and jump operator takes the form $L_{\nu\mu}=\sqrt{W_{\nu\mu}}|\nu\rangle\langle\mu|$. However, in quantum dynamics, the degree of activity must account not only for jumps but also for smooth and continuous time evolution. Additional terms contribute to the overall activity by capturing the effects of continuous time evolution.

Nishiyama and Hasegawa derived the following expression \cite{nishiyamaExactSolutionQuantum2024}:
\begin{equation}\label{Nishiyama_QDA}
\begin{split}
    \mathcal{B}(\tau) =& \mathcal{A}(\tau) + 8 \int_0^\tau ds_1 \int_0^{s_1} ds_2 \, \mathrm{Re}\Big(\mathrm{Tr}_S[H_\mathrm{eff}^\dagger \check{H}(s_1 - s_2) \\
    & \times \rho(s_2) ] \Big) - 4 \left( \int_0^\tau ds \, \mathrm{Tr}_S[H \rho(s)] \right)^2,
\end{split}
\end{equation}
where $\check{H}(t)\equiv e^{\mathcal{L}^\dagger t}H$ with $\mathcal{L}^\dag$ being the adjoint superoperator defined by
\begin{equation}\label{adjoint_lindblad}
    \dot{\mathcal{O}}=\mathcal{L}^\dag\mathcal{O}\equiv i[H, \mathcal{O}]+\sum_{z=1}^{N_c}L_z^\dag\mathcal{O}L_z-\frac{1}{2}\{L_z^\dag L_z, \mathcal{O}\},
\end{equation}
where $\mathcal{O}$ is the operator. \EEref{adjoint_lindblad} corresponds to the time evolution in the Heisenberg picture. Considering the classical limit for \Eref{Nishiyama_QDA}, i.e., setting $H=0$, we find that $\mathcal{B}(\tau)=\mathcal{A}(\tau)$. This result signifies that the quantum dynamical activity reduced to the classical dynamical activity in the classical limit. Next, we consider the closed limit, i.e., setting $L_z=0$, we find the following relation:
\begin{equation}
    \mathcal{B}(\tau)=4\tau^2(\mathrm{Tr}_S[H^2\rho]-\mathrm{Tr}_S[H\rho]^2)=4\tau^2\mathrm{Var}(H).
\end{equation}
Note that for the isolated system, the expectation value of the energy is conserved, and $H$ commutes with $e^{iHt}$.  By substituing this result into \Eref{QSL}, we recover the Mandelstam and Tamm QSL [\Eref{mtbound}]. 

The Nakajima-Utsumi (NU) -type quantum dynamical activity [\Eref{Nakajima_QDA}] and the Nishiyama-Hasegawa (NH) -type quantum dynamical activity [\Eref{Nishiyama_QDA}] represent the exact same quantity but are expressed in different forms.

\subsection{Quantum Feedback Control} \label{qfb}
In this work, we consider the dynamics of a quantum system under Markovian feedback control, which is characterized by direct and real-time utilization of information obtained from the system; the measurement outcomes are immediately used to manipulate the system's subsequent evolution without any significant delay. As the source of this real-time information, we consider continuous measurements. To minimize disturbance of the quantum system, quantum feedback control often employs continuous measurement rather than projective measurement. Specifically, our work covers three paradigmatic types of continuous measurement: jump measurement, homodyne measurement, and Gaussian measurement.

Unraveling of Lindblad dynamics [\Eref{unraveling_lindblad}], which does not include feedback control, can be rewritten as 
\begin{equation}\label{unraveling_lindblad2}
    \rho_c(t+dt)=e^{\mathcal{H}dt}\frac{M_z\rho_c(t)M_z^\dag}{p_z(t)},
\end{equation}
which indicates that the unitary time evolution due to the Hamiltonian $H$ occurs after the measurement associated with $M_z$. In this formulation, $M_0$ is rewritten as $M_0=1-\frac{1}{2}\sum_{z=1}^{N_c}L_z^\dagger L_z dt$. The Lindblad equation can be obtained by taking the average of \Eref{unraveling_lindblad2}.

In order to perform feedback control using the result of jump measurement, we consider applying a control input proportional to the current $I(t)$ of the jump measurement as a unitary time evolution by the Hermitian operator $F$. Under this situation, we obtain the following unraveling:
\begin{equation}\label{jump_unraveling}
\rho_c(t+dt)=e^{\mathcal{H}dt}e^{I(t)\mathcal{F}dt}\frac{M_z\rho_c(t)M_z^\dag}{p_z(t)},
\end{equation}
where $\mathcal{F}\rho\equiv-i[F, \rho]$. This unraveling indicates that the feedback control input is applied after the measurement is performed. By averaging \Eref{jump_unraveling}, we can derive the following equation \cite{wisemanQuantumTheoryContinuous1994}:
\begin{equation}\label{jump_dynamics}
\begin{split}
    \frac{d\rho}{dt} = & \mathcal{L}_\mathrm{J} \rho \\
    = & \mathcal{H} \rho + \sum_{z=1}^{N_c} \left[ e^{\nu_z \mathcal{F}} (L_z \rho L_z^\dagger) - \frac{1}{2} L_z^\dagger L_z \rho - \frac{1}{2} \rho L_z^\dagger L_z \right],
\end{split}
\end{equation}
which describes the dynamics under feedback control by jump measurements.

Similarly, considering feedback control by homodyne measurement, we apply a unitary time evolution proportional to the current $I_\mathrm{hom}(t)$, leading to
\begin{equation}\label{homodyne_unraveling}
\rho_c(t+dt)=e^{\mathcal{H}dt}e^{I_\mathrm{hom}(t)\mathcal{F}dt}\frac{M_z\rho_c(t)M^\dag_z}{p_z(t)}.
\end{equation}
Note that the Kraus operators and Hamiltonian for homodyne measurement are
\begin{equation}\label{homodyne_kraus_op}
\begin{split}
    &\mathcal{H}\rho=-i\left[H-\frac{i}{2}\sum_z (\alpha_z^* L_z-\alpha_zL_z^\dagger), \rho\right]\\
    &M_0=\mathbb{I}-\frac{1}{2}\sum_z\left(L^\dagger_z L_z+\alpha_zL_z^\dagger+\alpha^* L_z+|\alpha_z|^2\right)dt\\
    &M_z= \sqrt{dt}(L_z+\alpha_z), 
\end{split}
\end{equation}
because of \Eref{homodyne_transform}, where $|\alpha_z|\rightarrow\infty$. In the following, we set $\nu_z=1$ when considering feedback control by homodyne measurement. We can recover other values of $\nu_z$ by the substitution $F\rightarrow \nu_z F$. By averaging \Eref{homodyne_unraveling}, we can derive the following equation \cite{wisemanQuantumTheoryOptical1993}:
\begin{equation}\label{real_homodyne_dynamics}
\begin{split}
\frac{d\rho}{dt} = & \mathcal{L}_\mathrm{H} \rho \\
= & \mathcal{H} \rho + \sum_{z=1}^{N_c}\Big[\mathcal{D}[L_z]\rho \\
&+ \mathcal{F}(e^{-i\phi_z}L_z\rho+e^{i\phi_z}\rho L_z^\dagger) + \frac{1}{2} \mathcal{F}^2 \rho\Big].
\end{split}
\end{equation}
These master equations for homodyne measurement are known as Wisemen-Milburn equation. 

Likewise, for the dynamics under feedback control by Gaussian measurement, applying a similar procedure with $M_z$ in \Eref{gaussian_kraus} and $I_{\mathrm{gau}}(t)$ yields the following equation \cite{annby-doc}:
\begin{equation}\label{homodyne_dynamics}
\begin{split}
\frac{d\rho}{dt} = & \mathcal{L}_\mathrm{G} \rho \\
= & \mathcal{H} \rho + \lambda \mathcal{D}[Y] \rho + \frac{1}{2} \mathcal{F}\{Y, \rho\} + \frac{1}{8 \lambda} \mathcal{F}^2 \rho.
\end{split}
\end{equation}

Notably, the formulation of quantum feedback control adopted here does not require complex processing such as real-time state estimation or adaptive adjustments of feedback. Consequently, it is conceptually simple and more straightforward to implement in practical applications.

\section{Results}

Reference \cite{hasegawaQuantumThermodynamicUncertainty2023}, an unpublished preprint by one of the authors, presents a preliminary version of the quantum TUR under feedback control, which uses jump measurement results and Gaussian measurement results, based on cMPS method. However, the quantum dynamical activity, which is used as the cost in the quantum TUR, is defined solely using quantum Fisher information, and its exact analytical representation remains unclear. To address this issue, our study provides a more comprehensive and rigorous formulation. We construct the cMPS with reference to Ref. \cite{hasegawaQuantumThermodynamicUncertainty2023}, and follow it to derive quantum TUR under feedback control. Most importantly, using the cMPS, we analytically derive the exact quantum dynamical activity under feedback control based on Nakajima-Utsumi and Nishiyama-Hasegawa method \cite{nakajimaSymmetriclogarithmicderivativeFisherInformation2023, nishiyamaExactSolutionQuantum2024}. These representations express the quantum dynamical activity using only information of the system, which allows for a clear physical interpretation. Reference \cite{hasegawaQuantumThermodynamicUncertainty2023} is limited to jump and Gaussian measurements, but here we also consider homodyne measurements. Furthermore, we can derive two additional fundamental trade-off relations under feedback control. First, we derive QSL under feedback control. Our framework enables us to clarify the inherent duality between the quantum TUR and the QSL under feedback control. Second, we derive a more general quantum TUR by utilizing concentration inequalities under feedback control, which takes a form similar to \Eref{concentration_QTUR}.

\subsection{Jump Measurement}
We consider feedback control using jump measurement results. From \Eref{jump_unraveling}, Kraus representation is given by
\begin{equation}\label{jump_Kraus}
    \rho(t+dt)=\sum_z U_z M_z \rho(t) M_z^\dag U_z^\dag,
\end{equation}
where $U_z\equiv e^{-iHdt}e^{-i\nu_z F dt}$. From \Eref{jump_Kraus}, MPS can be defined as
\begin{equation}\label{jump_MPS}
    |\Psi(\tau)\rangle=\sum_{\bm{z}}U_{z_{N_\tau-1}}M_{z_{N_\tau-1}}\cdots U_{z_0}M_{z_0}|\psi(0)\rangle\otimes|\bm{z}\rangle,
\end{equation}
where $\bm{z}\equiv[z_{N_\tau-1}, \cdots z_{0}]$. Since the dynamics is given by \Eref{jump_dynamics}, the time scaling of $(1+\theta)$ is obtained by the following parametrization:
\begin{equation}\label{jump_scaling}
    H(\theta)=(1+\theta)H, L_z(\theta)=\sqrt{1+\theta}L_z, F(\theta)=F.
\end{equation}
From the structure of \Eref{jump_dynamics}, $F(\theta)$ is independent of $\theta$ to yield the desired time scaling. Defining $\rho^{\theta, \phi}(\tau)$ in the same way as in \Eref{two_rho}, $\rho^{\theta, \phi}(\tau)$ follows the two-sided version of \Eref{jump_dynamics}:
\begin{equation}\label{jump_two_sided}
\begin{split}
    \frac{d\rho^{\theta, \phi}}{dt} =&\mathcal{L}_\mathrm{J}^{\theta, \phi}\rho^{\theta, \phi} \\
    =&\mathcal{H}(\theta, \phi)\rho^{\theta, \phi} + \sum_{z} \Big[ e^{\nu_z \mathcal{F}(\theta, \phi)} (L_z(\theta)\rho^{\theta, \phi} L_z^\dagger(\phi))\\ 
&- \frac{1}{2}L_z^\dagger(\theta)L_z(\theta)\rho^{\theta, \phi} - \frac{1}{2}\rho^{\theta, \phi} L_z^\dagger(\phi)L_z(\phi) \Big],
\end{split}
\end{equation}
where $\mathcal{F}(\theta, \phi)\rho^{\theta, \phi}\equiv-i[F(\theta)\rho^{\theta, \phi}-\rho^{\theta, \phi}F(\phi)]$.
Under the parameterization of \Eref{jump_scaling}, we can define quantum dynamical activity under feedback control using jump measurement results $\mathcal{B}_\mathrm{jmp}^\mathrm{fb}(\tau)$ as follows:
\begin{equation}
    \mathcal{B}_\mathrm{jmp}^\mathrm{fb}(\tau)\equiv \mathcal{I}_\mathrm{jmp}^\mathrm{fb}(\theta)|_{\theta=0}
\end{equation}
where $\mathcal{I}_\mathrm{jmp}^\mathrm{fb}(\theta)$ is the quantum Fisher information with MPS in \Eref{jump_MPS}. Here, by using the Cram\'er-Rao inequality [\Eref{cramer}], we can derive the following quantum TUR:
\begin{equation}\label{jump_tur_notsteady}
    \frac{\mathrm{Var}[N(\tau)]}{\tau^2(\partial_\tau\langle N(\tau) \rangle)^2}\geq \frac{1}{\mathcal{B}_{\mathrm{jmp}}^{\mathrm{fb}}(\tau)}.
\end{equation}
Under the steady state condition,  we can obtain
\begin{equation}\label{jump_tur}
    \frac{\mathrm{Var}[N(\tau)]}{\langle N(\tau) \rangle^2}\geq \frac{1}{\mathcal{B}_{\mathrm{jmp}}^{\mathrm{fb}}(\tau)}.
\end{equation}
These results demonstrate a fundamental trade-off between precision and cost under feedback control by jump measurement, dictating that a smaller variance of the observable necessitates larger cost $\mathcal{B}_{\mathrm{jmp}}^{\mathrm{fb}}(\tau)$. This cost $\mathcal{B}_{\mathrm{jmp}}^{\mathrm{fb}}(\tau)$ also accounts for the contribution from feedback.

Then, we derive exact expression of $\mathcal{B}_{\mathrm{jmp}}^{\mathrm{fb}}(\tau)$ based on NU-type of quantum dynamical activity [\Eref{Nakajima_QDA}] and NH-type of quantum dynamical activity [\Eref{Nishiyama_QDA}]. We find that NU-type of quantum dynamical activity under feedback control by jump measurement is
\begin{equation}\label{Nakajima_jump_QDA}
    \mathcal{B}_{\mathrm{jmp}}^{\mathrm{fb}}(\tau) =\mathcal{A}(\tau) + 4(I_{\mathrm{J}1} + I_{\mathrm{J}2}) - 4 \left( \int_0^\tau ds \, \mathrm{Tr}_S[H\rho(s)] \right)^2, 
\end{equation}
where 
\begin{equation}
I_{\mathrm{J}1} \equiv \int_0^\tau ds_1 \int_0^{s_1} ds_2 \, \mathrm{Tr}_S\left[\mathcal{K}_{\mathrm{J}2} e^{\mathcal{L}_{\mathrm{J}}(s_1 - s_2)} (\mathcal{K}_{\mathrm{J}1} \rho(s_2))\right],
\end{equation}
\begin{equation}
I_{\mathrm{J}2} \equiv \int_0^\tau ds_1 \int_0^{s_1} ds_2 \, \mathrm{Tr}_S\left[\mathcal{K}_{\mathrm{J}1} e^{\mathcal{L}_{\mathrm{J}}(s_1 - s_2)} (\mathcal{K}_{\mathrm{J}2} \rho(s_2))\right],
\end{equation}
with
\begin{equation}\label{k_j1}
    \mathcal{K}_{\mathrm{J}1} \bullet \equiv -iH_\mathrm{eff} \bullet + \frac{1}{2} \sum_z e^{\nu_z \mathcal{F}} (L_z \bullet L_z^\dagger),
\end{equation}
\begin{equation}\label{k_j2}
    \mathcal{K}_{\mathrm{J}2} \bullet \equiv i \bullet H_\mathrm{eff}^\dagger + \frac{1}{2} \sum_z e^{\nu_z \mathcal{F}} (L_z \bullet L_z^\dag).
\end{equation}
We find that NH-type of quantum dynamical activity under feedback control by jump measurement is
\begin{equation}\label{Nishiyama_jump_QDA}
\begin{split}
    \mathcal{B}_{\mathrm{jmp}}^{\mathrm{fb}}(\tau) =& \mathcal{A}(\tau) + 8 \int_0^\tau ds_1 \int_0^{s_1} ds_2 \, \mathrm{Re}\Big(\mathrm{Tr}_S[H_\mathrm{eff}^\dagger \\
    &\times\breve{H}(s_1 - s_2) \rho(s_2) ] \Big) - 4 \left( \int_0^\tau ds \, \mathrm{Tr}_S[H \rho(s)] \right)^2,
\end{split}
\end{equation}
where $\breve{H}(t)\equiv e^{\mathcal{L}_{\mathrm{J}}^\dagger t}H$ with $\mathcal{L}_{\mathrm{J}}^\dag$ being the adjoint superoperator corresponding to \Eref{jump_dynamics} defined by
\begin{equation}\label{jump_adjoint}
    \dot{\mathcal{O}}=\mathcal{L}_{\mathrm{J}}^\dag\mathcal{O}\equiv i[H, \mathcal{O}]+\sum_{z=1}^{N_c}L_z^\dag e^{\nu_z \mathcal{F}^\dagger}(\mathcal{O})L_z-\frac{1}{2}\{L_z^\dag L_z, \mathcal{O}\}.
\end{equation}
A detailed derivation of \Eref{Nakajima_jump_QDA} - (\ref{jump_adjoint}) is shown in Appendix \ref{jump_derivation}. When there is no feedback, i.e., $F=0$, both NU- and NH-types of quantum dynamical activity $\mathcal{B}_{\mathrm{jmp}}^{\mathrm{fb}}(\tau)$ [\Eref{Nakajima_jump_QDA}, \Eref{Nishiyama_jump_QDA}] are equal to quantum dynamical activity for Lindblad dynamics $\mathcal{B}(\tau)$ [\Eref{Nakajima_QDA}, \Eref{Nishiyama_QDA}]. This means that, in the absence of feedback, quantum TUR in \Eref{jump_tur_notsteady} and \Eref{jump_tur} reduces to the quantum TUR for continuous measurement [\Eref{lindblad_jump_tur}, \Eref{lindblad_jump_tur_steady}].

We can also obtain QSL under feedback control by jump measurement by following an derivation analogous to the QSL for Lindblad dynamics [\Eref{QSL}], employing MPS [\Eref{jump_MPS}] and $\mathcal{B}_{\mathrm{jmp}}^{\mathrm{fb}}(\tau)$ as follows:
\begin{equation}\label{jump_qsl}
    \mathcal{L}_D(\rho(0), \rho(\tau))\leq\frac{1}{2}\int_0^\tau dt \frac{\sqrt{\mathcal{B}_{\mathrm{jmp}}^{\mathrm{fb}}(t)}}{t}.
\end{equation}
This indicate that a greater cost $\mathcal{B}_{\mathrm{jmp}}^{\mathrm{fb}}(\tau)$ is necessary for achieving a higher speed under feedback control by jump measurement.

Quantum TUR from concentration inequality under feedback control by jump measurement can also be derived by following a derivation of the one for Lindblad dynamics [\Eref{concentration_QTUR}] as follows:
\begin{equation}\label{jump_fb_concent}
    \frac{\|N^\circ(\tau)\|_p}{\|N^\circ(\tau)\|_1}\geq \sin{\left[\frac{1}{2}\int_0^\tau dt\frac{\sqrt{\mathcal{B}_{\mathrm{jmp}}^{\mathrm{fb}}(t)}}{t}\right]}^{-\frac{2(p-1)}{p}},
\end{equation}
which holds for $0\leq(1/2)\int_0^\tau dt\sqrt{\mathcal{B}_{\mathrm{jmp}}^{\mathrm{fb}}(t)}/t\leq\pi/2$. A more general trade-off relation between precision and cost under feedback control is obtained with $\mathcal{B}_{\mathrm{jmp}}^{\mathrm{fb}}(\tau)$ serving as the cost. In the limit of no feedback, i.e., for $F=0$, \Eref{jump_qsl} and \Eref{jump_fb_concent} reduce to their corresponding trade-off relations for standard Lindblad dynamics [\Eref{QSL}, \Eref{concentration_QTUR}].

The foregoing analysis thus establishes quantum dynamical activity under feedback control by jump measurement $\mathcal{B}_{\mathrm{jmp}}^{\mathrm{fb}}(\tau)$ as a crucial quantity that constrains both the speed and precision of the dynamics under feedback control by jump measurement. To investigate the feedback contribution to $\mathcal{B}_{\mathrm{jmp}}^{\mathrm{fb}}(\tau)$, we employ Taylor expansion. We decompose $\mathcal{B}_\mathrm{jmp}^\mathrm{fb}(\tau)$ into the quantum dynamical activity in the absence of feedback control $\mathcal{B}(\tau)$ and a feedback-dependent term. Under the assumption of $\tau \ll 1$ and $\nu_z \ll 1$, $\mathcal{B}_\mathrm{jmp}^\mathrm{fb}(\tau)$ becomes as follows:
\begin{equation}\label{jump_taylor}
\begin{split}
    \mathcal{B}_{\mathrm{jmp}}^{\mathrm{fb}}(\tau) \simeq &\mathcal{B}(\tau)+8 \int_0^\tau ds_1\int_0^{s_1} ds_2 \mathrm{Re}\Big\{\mathrm{Tr}_S\big[H_\mathrm{eff}^\dagger \\
    &\times\Big(\sum_z i\nu_zL_z^\dagger[F, H] L_z\Big)(s_1 - s_2) \rho(s_2) \big] \Big\}.
\end{split}
\end{equation}
The derivation is shown in the Appendix \ref{jump_taylor_derivation}. Under this assumption, the commutator between the Hamiltonian $H$ and the Hermitian operator $F$ used for the feedback plays a crucial role in determining how the feedback control affects the precision and speed of the dynamics. This means that if the feedback operator $F$ is chosen to be proportional to the Hamiltonian of the system $H$, feedback contribution does not lead to improvement in the limits on precision and speed. Moreover, under this assumption, stronger feedback, as parameterized by a larger $\nu_z$, directly improves the fundamental limits on precision and speed.

\subsection{Homodyne Measurement}
Next, we consider feedback control using homodyne measurement results. Basically, we derive quantum TUR, analytical solution of quantum dynamical activity, and QSL in the same way in the case of feedback control by jump measurement. MPS for homodyne measurement based feedback control can also be constructed in the same form as \Eref{jump_MPS} by employing the Kraus operators $M_z$ in \Eref{homodyne_kraus_op} and $U_z\equiv e^{-iHdt-\frac{1}{2}\sum_z (\alpha_z^* L_z-\alpha_z L_z^\dagger)dt}e^{-iI_{\mathrm{hom}}(t) F dt}$. To obtain the time scaling of $(1+\theta)$ for the dynamics in \Eref{real_homodyne_dynamics}, we suppose the following parametrization:
\begin{equation}\label{homodyne_scaling}
    H(\theta)=(1+\theta)H, L_z(\theta)=\sqrt{1+\theta}L_z, F(\theta)=\sqrt{1+\theta}F.
\end{equation}
Note that $F(\theta)$ is dependent on $\theta$ here, which gives rise to desired time scaling because of the structure of \Eref{real_homodyne_dynamics}. Two-sided version of \Eref{real_homodyne_dynamics} becomes
\begin{equation}\label{homodyne_two_sided}
\begin{split}
\frac{d\rho^{\theta, \phi}}{dt} =&\mathcal{L}_\mathrm{H}^{\theta, \phi}\rho^{\theta, \phi} \\
=&\mathcal{H}(\theta, \phi) \rho^{\theta, \phi} + \sum_z L_z(\theta)\rho L_z^\dagger(\phi) \\
 &- \frac{1}{2}(\rho^{\theta, \phi} L_z^\dagger(\phi) L_z(\phi)+ L_z^\dagger(\theta) L_z(\theta) \rho^{\theta, \phi})  \\
&+ \mathcal{F}(\theta, \phi)(e^{-i\phi_z}L_z(\theta)\rho^{\theta, \phi} + e^{i\phi_z}\rho^{\theta, \phi}L_z^\dagger(\phi)) \\
&+ \frac{1}{2} \mathcal{F}(\theta, \phi)^2 \rho^{\theta, \phi}.
\end{split}
\end{equation}
We define quantum dynamical activity under feedback control by homodyne measurement $\mathcal{B}_\mathrm{hom}^\mathrm{fb}(\tau)$ as follows:
\begin{equation}
    \mathcal{B}_\mathrm{hom}^\mathrm{fb}(\tau)\equiv \mathcal{I}_\mathrm{hom}^\mathrm{fb}(\theta)|_{\theta=0}
\end{equation}
where $\mathcal{I}_\mathrm{hom}^\mathrm{fb}(\theta)$ is the quantum Fisher information with MPS for homodyne measurement based feedback control. From the Cram\'er-Rao inequality [\Eref{cramer}], we obtain the following quantum TUR under the steady state condition:
\begin{equation}\label{homodyne_tur}
    \frac{\mathrm{Var}[Z(\tau)]}{\langle Z(\tau) \rangle^2}\geq \frac{1}{4\mathcal{B}_{\mathrm{hom}}^{\mathrm{fb}}(\tau)}.
\end{equation}
This represents the trade-off relation between the precision of an observable and the cost $\mathcal{B}_\mathrm{hom}^\mathrm{fb}(\tau)$ under feedback control by homodyne measurement. Similar to the case of only continuous measurement [\Eref{hom_tur}], this relation has a coefficient of 4 before the cost term.

Next, we derive the analytical solution of $\mathcal{B}_{\mathrm{hom}}^{\mathrm{fb}}(\tau)$. We define effective Hamiltonian in Wiseman-Milburn equation for homodyne measurement as
\begin{equation}
    H_{\mathrm{eff}}^{\mathrm{h}}\equiv H+\sum_z\left(-\frac{i}{2}L_z^\dagger L_z+e^{-i\phi_z}FL_z-\frac{i}{2}F^2\right).
\end{equation}
NU-type of quantum dynamical activity under feedback control by homodyne measurement is
\begin{equation}\label{Nakajima_homodyne_QDA}
\begin{split}
    \mathcal{B}_{\mathrm{hom}}^{\mathrm{fb}}(\tau)=&\mathcal{A}_{\mathrm{hom}}^{\mathrm{fb}}(\tau) + 4(I_{\mathrm{H}1}+ I_{\mathrm{H}2}) \\
    &- 4 \Big\{ \int_0^\tau ds \mathrm{Tr}_S\big[H\rho(s)\\
    &+\frac{1}{2}\sum_z(e^{i\phi_z}F\rho(s)L_z^\dagger+e^{-i\phi_z}FL_z\rho(s))\big]\Big\}^2, 
\end{split}
\end{equation}
where 
\begin{equation}
\begin{split}
\mathcal{A}_{\mathrm{hom}}^{\mathrm{fb}}(\tau)\equiv& \mathcal{A}(\tau)+ \int_0^\tau dt \sum_z\mathrm{Tr}_S\Big[ie^{-i\phi_z}L_z\rho(t)F\\
&-ie^{i\phi_z}F\rho(t)L_z^\dagger+F\rho(t)F\Big],
\end{split}
\end{equation}
\begin{equation}
I_{\mathrm{H}1} \equiv \int_0^\tau ds_1 \int_0^{s_1} ds_2 \, \mathrm{Tr}_S\left[\mathcal{K}_{\mathrm{H}2} e^{\mathcal{L}_{\mathrm{H}}(s_1 - s_2)} \mathcal{K}_{\mathrm{H}1} \rho(s_2)\right],
\end{equation}
\begin{equation}
I_{\mathrm{H}2} \equiv \int_0^\tau ds_1 \int_0^{s_1} ds_2 \, \mathrm{Tr}_S\left[\mathcal{K}_{\mathrm{H}1} e^{\mathcal{L}_{\mathrm{H}}(s_1 - s_2)} \mathcal{K}_{\mathrm{H}2} \rho(s_2)\right],
\end{equation}
with
\begin{equation}\label{k_h1}
\begin{split}
    \mathcal{K}_{\mathrm{H}1} \bullet \equiv& -iH_\mathrm{eff}^{\mathrm{h}} \bullet +\sum_z \Big[\frac{1}{2}L_z\bullet L_z^\dagger\\
    &-\frac{i}{2}e^{i\phi_z}F\bullet L_z^\dagger +\frac{i}{2}e^{-i\phi_z}L_z\bullet F+\frac{1}{2}F\bullet F\Big],
\end{split}
\end{equation}
\begin{equation}\label{k_h2}
\begin{split}
    \mathcal{K}_{\mathrm{H}2} \bullet \equiv& i\bullet H_\mathrm{eff}^{\mathrm{h}\dagger}  +\sum_z \Big[\frac{1}{2}L_z\bullet L_z^\dagger\\
    &-\frac{i}{2}e^{i\phi_z}F\bullet L_z^\dagger +\frac{i}{2}e^{-i\phi_z}L_z\bullet F+\frac{1}{2}F\bullet F\Big].
\end{split}
\end{equation}
$\mathcal{A}_{\mathrm{hom}}^{\mathrm{fb}}(\tau)$ can be divided into a term $\mathcal{A}(\tau)$ corresponding to classical dynamical activity and other terms corresponding to feedback control contribution.
NH-type of quantum dynamical activity under feedback control by homodyne measurement is
\begin{equation}\label{Nishiyama_homodyne_QDA}
\begin{split}
    &\mathcal{B}_{\mathrm{hom}}^{\mathrm{fb}}(\tau) \\
    &= \mathcal{A}_{\mathrm{hom}}^{\mathrm{fb}}(\tau) \\
    &+4 \int_0^\tau ds_1 \int_0^{s_1} ds_2 \mathrm{Re}\Big( \mathrm{Tr}_S\Big[2H_\mathrm{eff}^{\mathrm{h}\dag}\widetilde{H}(s_1-s_2)\rho(s_2)\\
    &+\sum_z H_\mathrm{eff}^{\mathrm{h}\dag}(e^{i\phi_z}\widetilde{L_z^\dagger F}(s_1-s_2)+e^{-i\phi_z}\widetilde{FL_z}(s_1-s_2))\rho(s_2)\big]\Big) \\
    &- 4 \Big\{ \int_0^\tau ds \mathrm{Tr}_S\big[H\rho(s)\\
    &+\frac{1}{2}\sum_z(e^{i\phi_z}F\rho(s)L_z^\dagger+e^{-i\phi_z}FL_z\rho(s))\Big]\Big\}^2, 
\end{split}
\end{equation}
where $\widetilde{\bullet}(t)\equiv e^{\mathcal{L}_{\mathrm{H}}^\dagger t}\bullet$ with $\mathcal{L}_{\mathrm{H}}^\dag$ being the adjoint superoperator corresponding to \Eref{real_homodyne_dynamics} defined by
\begin{equation}\label{adjoint_homodyne}
\begin{split}
    \dot{\mathcal{O}}=&\mathcal{L}_\mathrm{H}^\dag \mathcal{O} \\
    =& i[H, \mathcal{O}]+ \sum_z \Big[L_z^\dagger\mathcal{O}L_z-\frac{1}{2}\{L_z^\dagger L_z, \mathcal{O}\}\\
    &+(e^{-i\phi_z}\mathcal{F}^\dag (\mathcal{O}) L_z+e^{i\phi_z}L_z^\dagger\mathcal{F}^\dag (\mathcal{O}))+\frac{(\mathcal{F}^{\dag})^2}{2}\mathcal{O}\Big].
\end{split}
\end{equation}
A detailed derivation of \Eref{Nakajima_homodyne_QDA} - (\ref{adjoint_homodyne}) is shown in Appendix \ref{homodyne_derivation}. When there is no feedback, i.e., $F=0$, $\mathcal{B}_{\mathrm{hom}}^{\mathrm{fb}}(\tau)$ [\Eref{Nakajima_homodyne_QDA}, \Eref{Nishiyama_homodyne_QDA}] are equal to quantum dynamical activity for Lindblad dynamics $\mathcal{B}(\tau)$ [\Eref{Nakajima_QDA}, \Eref{Nishiyama_QDA}]. Thus, for $F=0$, quantum TUR in \Eref{homodyne_tur} reduces to the quantum TUR for purely homodyne measurement [\Eref{hom_tur}].

QSL under feedback control by homodyne measurement becomes
\begin{equation}\label{homodyne_qsl}
    \mathcal{L}_D(\rho(0), \rho(\tau))\leq\frac{1}{2}\int_0^\tau dt \frac{\sqrt{\mathcal{B}_{\mathrm{hom}}^{\mathrm{fb}}(t)}}{t}.
\end{equation}
This relation indicates that accelerating the dynamics under feedback control by homodyne measurement necessitates a large cost $\mathcal{B}_{\mathrm{hom}}^{\mathrm{fb}}(\tau)$. Furthermore, in the absence of feedback, i.e., $F=0$, this relation correctly reduces to the QSL for Lindblad dynamics [\Eref{QSL}]. 

As has been shown, the quantity $\mathcal{B}_{\mathrm{hom}}^{\mathrm{fb}}(\tau)$, quantum dynamical activity under feedback control by homodyne measurement, is therefore a fundamental quantity that constrains both the precision and speed in this scenario.

\subsection{Gaussian Measurement}
Finally, we consider feedback control by Gaussian measurement. Gaussian measurement offers the advantage of a relatively straightforward physical interpretation compared to the other continuous measurement schemes. MPS for feedback control by Gaussian measurement can also be constructed in the same form as \Eref{jump_MPS} by employing the Kraus operators $M_z$ in \Eref{gaussian_kraus} and $U_z\equiv e^{-iHdt}e^{-iI_{\mathrm{gau}}(t) F dt}$. We introduce the time scaling factor $(1+\theta)$ for the dynamics in \Eref{homodyne_dynamics} by the following parametrization:
\begin{equation}\label{gaussian_scaling}
    H(\theta)=(1+\theta)H, Y(\theta)=\sqrt{1+\theta}Y, F(\theta)=\sqrt{1+\theta}F.
\end{equation}
The two-sided version of \Eref{homodyne_dynamics} becomes
\begin{equation}\label{gaussian_two_sided}
\begin{split}
\frac{d\rho^{\theta, \phi}}{dt} =&\mathcal{L}_\mathrm{G}^{\theta, \phi}\rho^{\theta, \phi} \\
=&\mathcal{H}(\theta, \phi) \rho^{\theta, \phi} + \lambda Y(\theta)\rho Y(\phi) - \frac{\lambda}{2} \rho^{\theta, \phi} Y(\phi)^2  \\
&-\frac{\lambda}{2} Y(\theta)^2 \rho^{\theta, \phi}+ \frac{1}{2} \mathcal{F}(\theta, \phi)(\rho^{\theta, \phi} Y(\phi)+ Y(\theta)\rho^{\theta, \phi}) \\
&+ \frac{1}{8 \lambda} \mathcal{F}(\theta, \phi)^2 \rho^{\theta, \phi}.
\end{split}
\end{equation}
We define quantum dynamical activity under feedback control by gaussian measurement $\mathcal{B}_\mathrm{gau}^\mathrm{fb}(\tau)$ as follows:
\begin{equation}
    \mathcal{B}_\mathrm{gau}^\mathrm{fb}(\tau)\equiv \mathcal{I}_\mathrm{gau}^\mathrm{fb}(\theta)|_{\theta=0}
\end{equation}
where $\mathcal{I}_\mathrm{gau}^\mathrm{fb}(\theta)$ is the quantum Fisher information with MPS for feedback control by Gaussian measurement. By using Cram\'er-Rao inequality [\Eref{cramer}], we can derive quantum TUR under the steady state condition:
\begin{equation}\label{gau_tur}
    \frac{\mathrm{Var}[Z(\tau)]}{\langle Z(\tau) \rangle^2}\geq \frac{1}{4\mathcal{B}_{\mathrm{gau}}^{\mathrm{fb}}(\tau)}.
\end{equation}
This result represents the trade-off relation between precision and cost $\mathcal{B}_\mathrm{gau}^\mathrm{fb}(\tau)$ under feedback control by Gaussian measurement.

Next, we derive exact expression of $\mathcal{B}_{\mathrm{gau}}^{\mathrm{fb}}(\tau)$. We define effective Hamiltonian for \Eref{homodyne_dynamics} as
\begin{equation}
    H_{\mathrm{eff}}^{\mathrm{g}}\equiv H-i\frac{\lambda}{2}Y^2+\frac{1}{2}FY-\frac{i}{8\lambda}F^2.
\end{equation}
NU-type of quantum dynamical activity under feedback control by Gaussian measurement is
\begin{equation}\label{Nakajima_gaussian_QDA}
\begin{split}
    \mathcal{B}_{\mathrm{gau}}^{\mathrm{fb}}(\tau)=&\mathcal{A}_{\mathrm{gau}}^{\mathrm{fb}}(\tau) + 4(I_{\mathrm{G}1}+ I_{\mathrm{G}2}) \\
    &- 4 \Big\{ \int_0^\tau ds \mathrm{Tr}_S\big[H\rho(s) \\
    &+\frac{1}{4}(F\rho(s)Y+FY\rho(s))\big]\Big\}^2, 
\end{split}
\end{equation}
where 
\begin{equation}
\begin{split}
\mathcal{A}_{\mathrm{gau}}^{\mathrm{fb}}(\tau)\equiv& \mathcal{A}(\tau)+\int_0^\tau dt \mathrm{Tr}_S\Big[\frac{i}{2}Y\rho(t)F\\
&-\frac{i}{2}F\rho(t)Y+\frac{1}{4\lambda}F\rho(t)F\Big],
\end{split}
\end{equation}
\begin{equation}
I_{\mathrm{G}1} \equiv \int_0^\tau ds_1 \int_0^{s_1} ds_2 \, \mathrm{Tr}_S\left[\mathcal{K}_{\mathrm{G}2} e^{\mathcal{L}_{\mathrm{G}}(s_1 - s_2)} \mathcal{K}_{\mathrm{G}1} \rho(s_2)\right],
\end{equation}
\begin{equation}
I_{\mathrm{G}2} \equiv \int_0^\tau ds_1 \int_0^{s_1} ds_2 \, \mathrm{Tr}_S\left[\mathcal{K}_{\mathrm{G}1} e^{\mathcal{L}_{\mathrm{G}}(s_1 - s_2)} \mathcal{K}_{\mathrm{G}2} \rho(s_2)\right],
\end{equation}
with
\begin{equation}\label{k_g1}
    \mathcal{K}_{\mathrm{G}1} \bullet \equiv -iH_\mathrm{eff}^{\mathrm{g}} \bullet + \frac{\lambda}{2}Y\bullet Y-\frac{i}{4}F\bullet Y +\frac{i}{4}Y\bullet F+\frac{1}{8\lambda}F\bullet F,
\end{equation}
\begin{equation}\label{k_g2}
    \mathcal{K}_{\mathrm{G}2} \bullet \equiv i\bullet H_\mathrm{eff}^{\mathrm{g}\dagger}  + \frac{\lambda}{2}Y\bullet Y-\frac{i}{4}F\bullet Y +\frac{i}{4}Y\bullet F+\frac{1}{8\lambda}F\bullet F.
\end{equation}
$\mathcal{A}_{\mathrm{gau}}^{\mathrm{fb}}(\tau)$ can be divided into a term $\mathcal{A}(\tau)$ which corresponds to classical dynamical activity, and other terms with feedback control contribution.
NH-type of quantum dynamical activity under feedback control by Gaussian measurement is
\begin{equation}\label{Nishiyama_gaussian_QDA}
\begin{split}
    &\mathcal{B}_{\mathrm{gau}}^{\mathrm{fb}}(\tau) \\
    &= \mathcal{A}_{\mathrm{gau}}^{\mathrm{fb}}(\tau) \\
    &+ \int_0^\tau ds_1 \int_0^{s_1} ds_2 \mathrm{Re}\Big(\mathrm{Tr}_S\big[8H_\mathrm{eff}^{\mathrm{g}\dag}\overline{H}(s_1-s_2)\rho(s_2)\\
    &+2H_\mathrm{eff}^{\mathrm{g}\dag}(\overline{YF}(s_1-s_2)+\overline{FY}(s_1-s_2))\rho(s_2)\big]\Big) \\
    &-4\Big(\int_0^\tau ds \mathrm{Tr}_S\big[H\rho(s)+\frac{1}{4}(F\rho(s)Y+FY\rho(s))\big]\Big)^2
\end{split}
\end{equation}
where $\overline{\bullet}(t)\equiv e^{\mathcal{L}_{\mathrm{G}}^\dagger t}\bullet$ with $\mathcal{L}_{\mathrm{G}}^\dag$ being the adjoint superoperator corresponding to \Eref{homodyne_dynamics} defined by
\begin{equation}\label{adjoint_gaussian}
\begin{split}
    \dot{\mathcal{O}}=&\mathcal{L}_G^\dag \mathcal{O} \\
    =& i[H, \mathcal{O}]+ \lambda Y\mathcal{O}Y-\frac{\lambda}{2}\{Y^2, \mathcal{O}\}\\
    &+\frac{1}{2}(\mathcal{F}^\dag (\mathcal{O}) Y+Y\mathcal{F}^\dag (\mathcal{O}))+\frac{(\mathcal{F}^{\dag})^2}{8\lambda}\mathcal{O}.
\end{split}
\end{equation}
A detailed derivation of \Eref{Nakajima_gaussian_QDA} - (\ref{adjoint_gaussian}) is shown in Appendix \ref{gaussian_derivation}. Under the condition $L=\sqrt{\lambda}Y$, $\phi_z=0$, and $\nu=1/(2\sqrt{\lambda})$, which are the correspondence between homodyne and Gaussian measurement, $\mathcal{B}_{\mathrm{hom}}^{\mathrm{fb}}(\tau)$ becomes identical to $\mathcal{B}_{\mathrm{gau}}^{\mathrm{fb}}(\tau)$ for both NU- and NH-type. When there is no feedback, i.e., $F=0$, $\mathcal{B}_{\mathrm{gau}}^{\mathrm{fb}}(\tau)$ [\Eref{Nakajima_gaussian_QDA}, \Eref{Nishiyama_gaussian_QDA}] are equal to quantum dynamical activity for Lindblad dynamics $\mathcal{B}(\tau)$ [\Eref{Nakajima_QDA}, \Eref{Nishiyama_QDA}]. Therefore, quantum TUR in \Eref{gau_tur} also reduces to \Eref{hom_tur} in the absence of feedback (i.e., $F=0$).

QSL under feedback control by Gaussian measurement becomes
\begin{equation}\label{gaussian_qsl}
    \mathcal{L}_D(\rho(0), \rho(\tau))\leq\frac{1}{2}\int_0^\tau dt \frac{\sqrt{\mathcal{B}_{\mathrm{gau}}^{\mathrm{fb}}(t)}}{t}.
\end{equation}
This result describes the trade-off relation under feedback control by Gaussian measurement, dictating that a higher speed requires substantial cost $\mathcal{B}_{\mathrm{gau}}^{\mathrm{fb}}(\tau)$. As with the other feedback schemes, this relation correctly reduces to the QSL for standard Lindblad dynamics when feedback is absent.

For feedback control by Gaussian measurement, $\mathcal{B}_{\mathrm{gau}}^{\mathrm{fb}}(\tau)$ is the fundamental quantity that constrains both precision and speed. To elucidate its role, we examine the exact representation of $\mathcal{B}_{\mathrm{gau}}^{\mathrm{fb}}(\tau)$, noting that $\lambda$ corresponds to the measurement strength. We find that as $\lambda$ increases, the influence of $\mathcal{B}_{\mathrm{gau}}^{\mathrm{fb}}(\tau)$ on the precision and speed limits is increasingly dominated by the contribution from the measurement operator $Y$, while the contribution from the feedback operator $F$ diminishes. Physically, this implies that the change in the quantum state induced by the measurement becomes more critical in determining the precision and speed than the change driven by the feedback input.

The derived results are summarized in Table \ref{result_table}.

\section{Numerical Simulations of Derived Trade-off Relations}
To verify our results, we apply the QSL and quantum TUR under feedback control by jump and homodyne measurement to concrete models under feedback control, and perform numerical simulations. Furthermore, we investigate how the presence of feedback control affects the speed and the precision.

\subsection{Two-Level Atom Driven by Classical Laser Field}\label{two-level}

As our first example, we consider a two-level atom driven by a classical laser field, which constitutes a dissipative Rabi oscillation model. The coherent evolution and dissipative dynamics of this system are described by the following Hamiltonian $H$ and jump operator $L$:
\begin{equation}
    H=\Delta|e\rangle\langle e| + \frac{\Omega}{2}(|e\rangle\langle g| + |g\rangle\langle e| ), L=\sqrt{\kappa}|g\rangle\langle e|,
\end{equation}
where $|e\rangle$ and $|g\rangle$ denote excited and ground states, respectively. The Hamiltonian models the interaction of the atom with the external laser field. Here, $\Omega$ is the Rabi frequency, and $\Delta$ is the detuning parameter defined as the difference between the laser's frequency and the atomic transition frequency. Furthermore, the jump operator $L$ describes an irreversible dissipative process, arising from the system's coupling to its environment, which induces quantum jumps from the excited state to the ground state via spontaneous emission with decay rate $\kappa$. A similar model leveraging Rabi oscillations has been experimentally implemented as a quantum clock to test quantum TUR \cite{heEffectMeasurementBackaction2023}. Therefore, we adopt this model as a well-suited testbed for examining the derived quantum thermodynamic trade-off relations.

For the feedback operator $F$, we employ the Pauli-X operator as a simple illustrative example:
\begin{equation}
    F = |e\rangle\langle g| + |g\rangle\langle e|.
\end{equation}
Indeed, it has been shown that for a pair of two-level atoms undergoing Rabi oscillations, Markovian feedback control employing the Pauli-X operator as $F$, based on jump or homodyne measurements, can enhance the entanglement between the atoms \cite{wangDynamicalCreationEntanglement2005, carvalhoControllingEntanglementDirect2008}.

\subsubsection{QSL}

\begin{figure}[t]
\includegraphics[width=60mm]{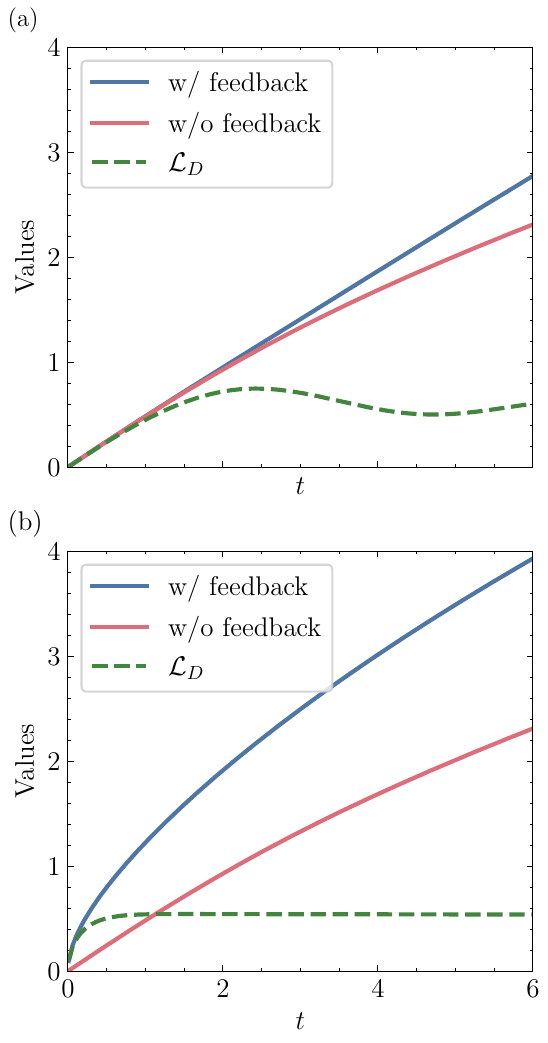}
\caption{\label{fig:fbqsl}Numerical simulation of QSL under feedback control. (a) and (b) show the cases of jump and homodyne measurement-based feedback control [\Eref{jump_qsl}, \Eref{homodyne_qsl}], respectively. The dashed lines denote the left-hand side (LHS) of the QSL: $\mathcal{L}_D(\rho(0), \rho(t))$. The blue solid lines represent the right-hand side (RHS) of the QSL. The red solid lines denote RHS of the QSL with quantum dynamical activity in the absence of feedback control. The parameters are set to $\Delta=1.0$, $\Omega=1.0$, $\kappa=0.5$, $\nu=1.0$ and $\phi=\pi/2$. 
} 
\end{figure}

Figures \ref{fig:fbqsl}(a) and (b) present numerical verifications of the QSL under feedback control by jump measurement  [\Eref{jump_qsl}] and homodyne measurement [\Eref{homodyne_qsl}], respectively. The blue solid lines represent the upper bound of the QSL [\Eref{jump_qsl}, \Eref{homodyne_qsl}], calculated using the quantum dynamical activity under feedback control. The dashed lines show the distance $\mathcal{L}_D(\rho(0), \rho(t))$ between the initial state and the state at time $t$, which is obtained by evolving the quantum state under the corresponding dynamics for jump [\Eref{jump_dynamics}] and homodyne measurements [\Eref{real_homodyne_dynamics}], respectively. In both cases, the dashed line is below the blue solid line, confirming that those QSL under feedback control are satisfied. Furthermore, to elucidate the role of the feedback contribution for speed, we compare these results with the case lacking feedback control. The red solid lines represent the right-hand side (RHS) of the QSL with quantum dynamical activity in the absence of feedback control. A clear separation is observed where the red solid line lies consistently below the blue solid line. This indicates that feedback control permits a larger achievable distance $\mathcal{L}_D(\rho(0), \rho(t))$ within the same time interval. In \Fref{fig:fbqsl}(b), there is a region at small $t$ where the red solid line is clearly below the dashed line, violating the QSL under feedback control. Consequently, this result shows that the presence of feedback control can effectively accelerate the quantum state evolution.

\subsubsection{Quantum TUR}

\begin{figure}[t]
\includegraphics[width=70mm]{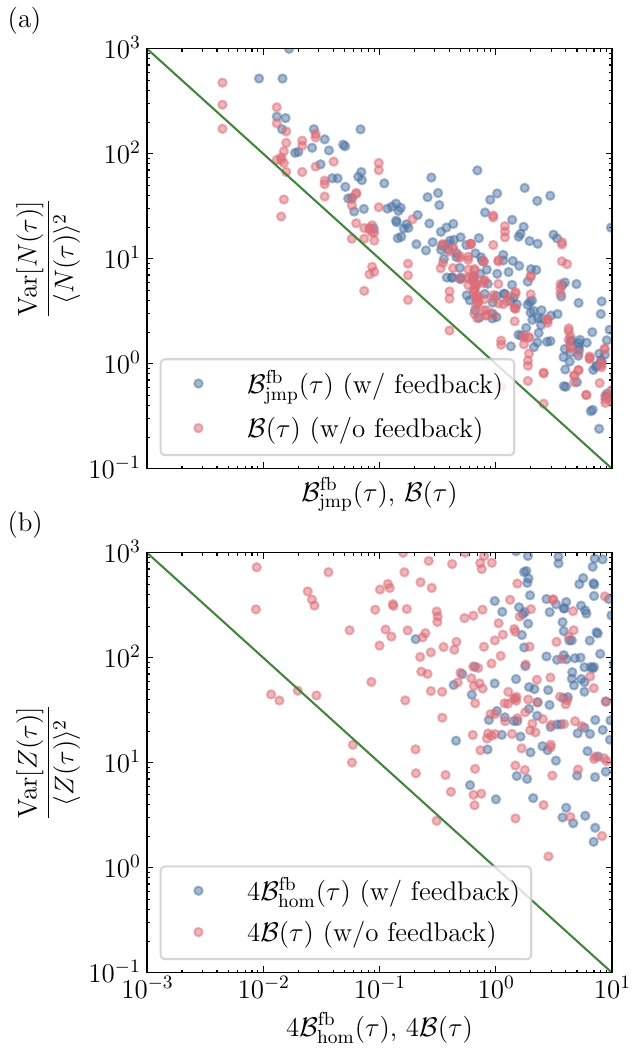}
\caption{\label{fig:fbtur}Numerical simulation of quantum TUR under feedback control. (a) and (b) show the cases of jump and homodyne measurement-based feedback control [\Eref{jump_tur}, \Eref{homodyne_tur}], respectively. Many random realizations are made and precision ($\mathrm{Var}[N(\tau)]/\langle N(\tau)\rangle^2$ and $\mathrm{Var}[Z(\tau)]/\langle Z(\tau)\rangle^2$) are plotted as a function of quantum dynamical activity. The blue and red dots use quantum dynamical activity with and without feedback, respectively, and the precision with feedback is plotted for them. Solid line denotes $1/\mathcal{B}_{\mathrm{jmp}}^{\mathrm{fb}}(\tau)$ and $1/4\mathcal{B}_{\mathrm{hom}}^{\mathrm{fb}}(\tau)$, respectively. Ranges for the parameters are $\Delta\in[0.1, 3.0]$, $\Omega\in[0.1, 3.0]$, $\kappa\in[0.1, 3.0]$, $\nu\in\{0.2, 0.4, 1.0\}$, $\phi\in[0, 2\pi]$, and $\tau\in[0.1, 3.0]$.
} 
\end{figure}

Next, we investigate quantum TUR. Figures \ref{fig:fbtur}(a) and (b) show numerical verification of the quantum TUR for steady state under feedback control by jump measurement [\Eref{jump_tur}] and homodyne measurement [\Eref{homodyne_tur}], respectively. To generate blue dots, we randomly sample system parameters from the ranges specified in the caption of \Fref{fig:fbtur}. For each parameter set, we simulate numerous quantum trajectories up to time $\tau$ to compute the mean and variance of the observable, thereby obtaining its precision (i.e., $\mathrm{Var}[N(\tau)]/\langle N(\tau)\rangle^2$, $\mathrm{Var}[Z(\tau)]/\langle Z(\tau)\rangle^2$). These precision values are then plotted as a function of the corresponding quantum dynamical activity under feedback control $\mathcal{B}_{\mathrm{jmp}}^{\mathrm{fb}}(\tau)$, $\mathcal{B}_{\mathrm{hom}}^{\mathrm{fb}}(\tau)$. The green solid line in each panel represents the theoretical lower bound of the quantum TUR, given by $1/\mathcal{B}_{\mathrm{jmp}}^{\mathrm{fb}}(\tau)$ and $1/4\mathcal{B}_{\mathrm{hom}}^{\mathrm{fb}}(\tau)$, respectively. As shown in both figures, all blue dots are above their respective solid lines, confirming that the quantum TUR under feedback control holds. Next, to investigate the impact of the feedback control for precision, we perform a comparative analysis. The red dots use the same precision values calculated from the feedback-controlled dynamics but are plotted against the quantum dynamical activity, $\mathcal{B}(\tau)$, of a system evolving without feedback. We observe that some red dots are below the solid line, indicating that quantum dynamical activity is smaller when there is no feedback. This result suggests that the feedback control not only modifies the system's evolution but also fundamentally increases the thermodynamic cost (i.e., quantum dynamical activity) required to achieve a certain level of measurement precision, thereby ensuring the validity of the quantum TUR.

\subsection{Quantum Error Correction}
Next, we focus on quantum error correction, which is one of the important applications of feedback control. More specifically, we consider quantum TUR for quantum error correction using continuous measurements, and perform numerical simulations. In contrast to typical error correction schemes based on discrete projective measurements and subsequent unitary feedback, the approach used here utilizes a continuous protocol. This scheme involves the continuous monitoring of errors and the application of immediate, Hamiltonian-based feedback upon their detection.

 In the field of quantum metrology \cite{giovannettiAdvancesQuantumMetrology2011}, where quantum systems are used as probes to estimate external field parameters, it is known that quantum error correction can restore quantum properties lost to noise, thereby improving sensitivity up to the Heisenberg limit \cite{durImprovedQuantumMetrology2014, arradIncreasingSensingResolution2014, kesslerQuantumErrorCorrection2014, ozeriHeisenbergLimitedMetrology2013, gefenParameterEstimationEfficient2016, undenQuantumMetrologyEnhanced2016, zhouAchievingHeisenbergLimit2018, huAchievingHeisenbergScaling2022, hongFeedbackControlQuantum2023, zhouAchievingMetrologicalLimits2024}. Drawing an analogy, we hypothesize that quantum error correction could similarly contribute to the improvement of precision in the context of the quantum TUR. Therefore, we perform numerical simulations of quantum error correction to investigate this possibility.

We consider the 2-qubit code with jump measurement proposed in \cite{ahnQuantumErrorCorrection2003}. This can be thought of as an operation to restore decoherence due to spontaneous photon emission. The code words are defined as follows:
\begin{equation}
\begin{split}
    &|0\rangle_L\equiv\frac{|00\rangle+|11\rangle}{2},\\
    &|1\rangle_L\equiv\frac{|01\rangle+|10\rangle}{2}.
\end{split}
\end{equation}
Since errors are represented as $|1\rangle\rightarrow|0\rangle$, the jump operators corresponding to errors for each qubit are
\begin{equation}
\begin{split}
    &L_1=\sqrt{\kappa_1}(\mathbb{X}+i\mathbb{Y})\otimes \mathbb{I},\\
    &L_2=\sqrt{\kappa_2}\mathbb{I}\otimes (\mathbb{X}+i\mathbb{Y}),
\end{split}
\end{equation}
where $\mathbb{X}, \mathbb{Y}, \mathbb{Z}$ are the Pauli operators. When an error by $L_1$ occurs, the code words become as follows:
\begin{equation}
\begin{split}
    &|0\rangle_L\rightarrow|01\rangle,\\
    &|1\rangle_L\rightarrow|00\rangle.
\end{split}
\end{equation}
Therefore, the feedback unitary to correct this error is
\begin{equation}
    U_1=\frac{\mathbb{X}\mathbb{I}+\mathbb{Z}\mathbb{X}}{\sqrt{2}}.
\end{equation}
Thus the Hermitian operator used for feedback is $F_1$ satisfying
\begin{equation}
    \exp(-iF_1dt)=U_1
\end{equation}
Similarly, when an error by $L_2$ occurs, the code word become
\begin{equation}
\begin{split}
    &|0\rangle_L\rightarrow|10\rangle,\\
    &|1\rangle_L\rightarrow|00\rangle,
\end{split}
\end{equation}
and the feedback unitary to correct this error is
\begin{equation}
    U_2=\frac{\mathbb{I}\mathbb{X}+\mathbb{X}\mathbb{Z}}{\sqrt{2}},
\end{equation}
and the correspinding Hermitian operator $F_2$ satisfy
\begin{equation}
    \exp(-iF_2dt)=U_2.
\end{equation}
These allow the error to be corrected when a jump can be detected by jump measurements. We need to further correct the dynamics during which jumps are not measured in the jump measurement. That is, we correct the dynamics corresponding to Kraus operator $M_0$ (defined in \Eref{Lindblad_Kraus}) that occurs between jumps. Here $M_0$ becomes
\begin{equation}
\begin{split}
    M_0&=\mathbb{I}-\frac{1}{2}L_1^\dagger L_1dt-\frac{1}{2}L_2^\dagger L_2dt-iHdt\\
    &=\mathbb{I}\mathbb{I}(1-(\kappa_1+\kappa_2)dt)+\kappa_1dt \mathbb{Z}\mathbb{I}+\kappa_2dt \mathbb{I}\mathbb{Z}-iHdt.
\end{split}
\end{equation}
When $H$ is defined as follows:
\begin{equation}
    H\equiv\kappa_1\mathbb{Y}\mathbb{X}+\kappa_2\mathbb{X}\mathbb{Y},
\end{equation}
we obtain
\begin{equation}
\begin{split}
    M_0=&\mathbb{I}\mathbb{I}(1-(\kappa_1+\kappa_2)dt)\\
    &+\kappa_1dt\mathbb{Z}\mathbb{I}(\mathbb{I}\mathbb{I}-\mathbb{X}\mathbb{X})+\kappa_2 dt\mathbb{I}\mathbb{Z}(\mathbb{I}\mathbb{I}-\mathbb{X}\mathbb{X}).
\end{split}
\end{equation}
Since $\mathbb{I}\mathbb{I}-\mathbb{X}\mathbb{X}$ works to annihilate the code words, time evolution due to $M_0$ is corrected.

\begin{figure}[t]
\includegraphics[width=70mm]{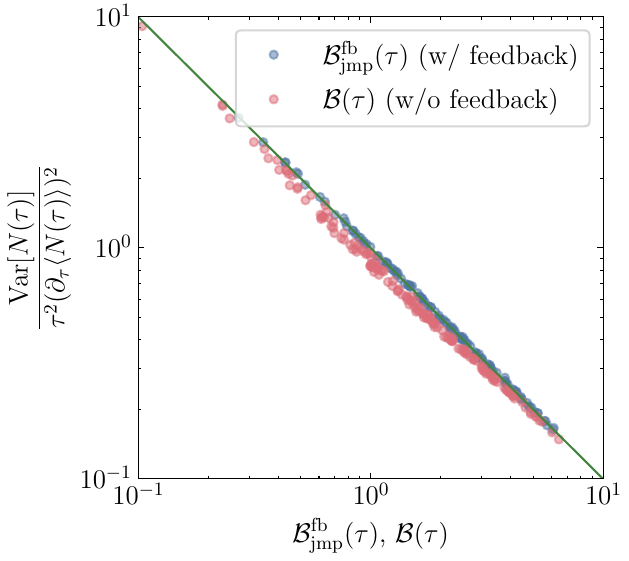}
\caption{\label{fig:fb_tur_qec}Numerical simulation of quantum TUR for quantum error correction. Many random realizations are made and precision $\mathrm{Var}[N(\tau)]/\tau^2(\partial_\tau\langle N(\tau)\rangle)^2$ are plotted as a function of quantum dynamical activity. The blue dots use quantum dynamical activity with feedback control and red dots use quantum dynamical activity without feedback control. Solid line denotes $1/\mathcal{B}_{\mathrm{jmp}}^{\mathrm{fb}}(\tau)$. Ranges for the parameters are $\kappa_1\in[0.1, 2.0]$, $\kappa_2\in[0.1, 2.0]$ and $\tau\in[0.1, 1.0]$.
} 
\end{figure}

We adapt such a setup of quantum error correction to quantum TUR under feedback control by jump measurement [\Eref{jump_tur_notsteady}] and perform numerical simulations. \FFref{fig:fb_tur_qec} shows the result of the numerical simulations. The blue dots plot precision [$\mathrm{Var}[N(\tau)]/\tau^2(\partial_\tau\langle N(\tau)\rangle)^2$] against quantum dynamical activity [$\mathcal{B}_{\mathrm{jmp}}^{\mathrm{fb}}(\tau)$] under feedback control by jump measurement. The solid line represents $1/\mathcal{B}_{\mathrm{jmp}}^{\mathrm{fb}}(\tau)$, and means lower bound for the quantum TUR [\Eref{jump_tur_notsteady}]. It can be seen that under this quantum error correction the precision of the continuous measurement is quite close to the theoretical limit. The red dots plot the precision with feedback [$\mathrm{Var}[N(\tau)]/\tau^2(\partial_\tau\langle N(\tau)\rangle)^2$] versus quantum dynamical activity without feedback control [$\mathcal{B}(\tau)$]. Without feedback control, quantum dynamical activity becomes small, which shows that the inequality is violated. This result also indicates that the presence of feedback control enhances precision.

\section{Detailed Analysis of Quantum Dynamical Activity}
To further elucidate the role of the cost term in the derived trade-off relations, we investigate the behavior of quantum dynamical activity under feedback control on the steady state of the simple two-level atom model introduced in Sec.~\ref{two-level}. We numerically analyze its dependence on time and feedback strength. We compare how the quantum dynamical activity behaves differently for feedback by jump and homodyne measurement.

\subsection{Time Dependence}
\begin{figure}[t]
\includegraphics[width=70mm]{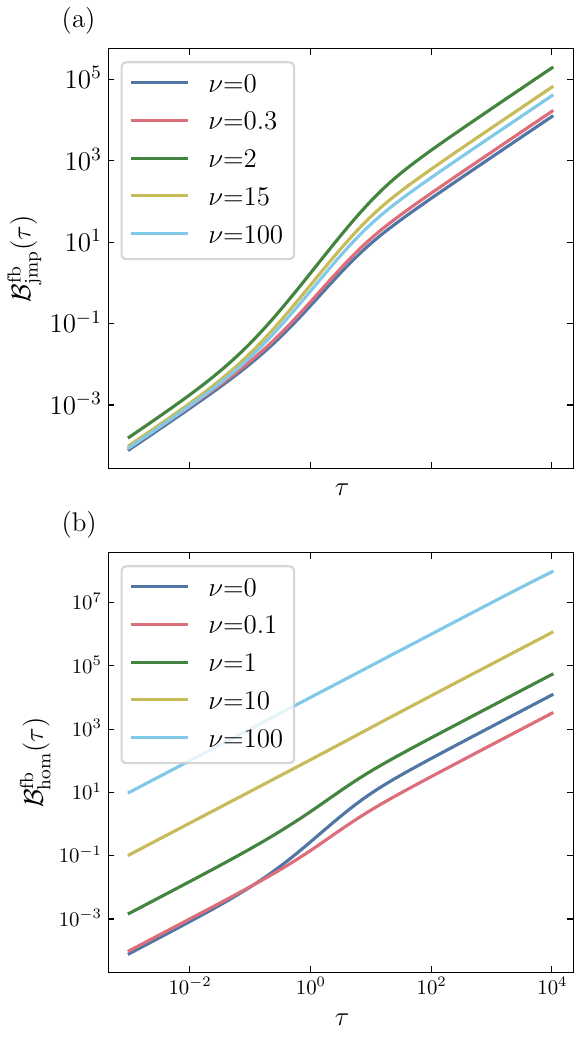}
\caption{\label{fig:qda_time} Numerical calculation of the time dependence of quantum dynamical activity under feedback control. (a) and (b) show the cases of feedback control by jump and homodyne measurement, respectively. The different curves represent the results for various feedback strengths, which correspond to the weights $ν$ in the output current $I(t)$ and $I_\mathrm{hom}(t)$, respectively. The parameters are set to $\Delta=1.0$, $\Omega=1.0$, $\kappa=0.5$, and $\phi_z=\pi/2$. 
} 
\end{figure}

First, we investigate the time dependence of the quantum dynamical activity under feedback control derived in the preceding sections. Figures \ref{fig:qda_time}(a) and (b) show the results of numerical calculations for the time evolution of the quantum dynamical activity under feedback control by jump $\mathcal{B}_\mathrm{jmp}^\mathrm{fb}(\tau)$ and homodyne measurements $\mathcal{B}_\mathrm{hom}^\mathrm{fb}(\tau)$, respectively. These calculations are performed on the two-level atom driven by a classical laser field introduced in Sec.~\ref{two-level}. The feedback strength is determined by the weight $\nu$ in the respective measurement current $I(t)$ [\Eref{jump_current}] and $I_\mathrm{hom}(t)$ [\Eref{homodyne_current}], and we plot the behavior for various values of $\nu$. Note that for feedback control by homodyne measurement, we consider the substitution $F\rightarrow \nu F$, as described in Sec.~\ref{qfb}. The case $\nu=0$ corresponds to the dynamics without feedback control, serving as a baseline for comparison.

Focusing on the case of feedback control by jump measurement, shown in \Fref{fig:qda_time}(a), we observe a non-trivial relationship: the quantum dynamical activity $\mathcal{B}_\mathrm{jmp}^\mathrm{fb}(\tau)$ does not necessarily increase with a larger feedback strength $\nu$. The underlying mechanism for this non-monotonic behavior will be elucidated in the following section [Sec.~\ref{strength_dependence}], where we analyze the dependence on feedback strength $\nu$ in detail. In contrast to the jump measurement case, feedback control by homodyne measurement exhibits a distinct behavior, as shown in \Fref{fig:qda_time}(b). With the exception of the regime of large $\tau$ and small $\nu$, the quantum dynamical activity generally increases with larger values of $\nu$.

The time dependence of quantum dynamical activity for Lindblad dynamics $\mathcal{B}(\tau)$ (i.e., in the absence of feedback control) has been previously investigated in Ref.~\cite{hasegawaQuantumThermodynamicUncertainty2020, nishiyamaExactSolutionQuantum2024}. It is known that $\mathcal{B}(\tau)$ grows linearly with $\tau$ in both the short-time ($\tau \ll 1$) and long-time ($\tau\rightarrow\infty$) regimes, while it exhibits superlinear scaling in the intermediate interval. This characteristic behavior is confirmed by our results for the $\nu=0$ case, as depicted in \Fref{fig:qda_time}. Furthermore, despite the presence of feedback, it is confirmed that $\mathcal{B}_\mathrm{jmp}^\mathrm{fb}(\tau)$ retains scaling properties similar to the no-feedback case $\mathcal{B}(\tau)$. In contrast, for $\mathcal{B}_\mathrm{hom}^\mathrm{fb}(\tau)$, we observe that the superlinear scaling, which is characteristic of the no-feedback dynamics in the intermediate time interval, is progressively suppressed as the feedback strength $\nu$ increases.

\begin{figure}[t]
\includegraphics[width=70mm]{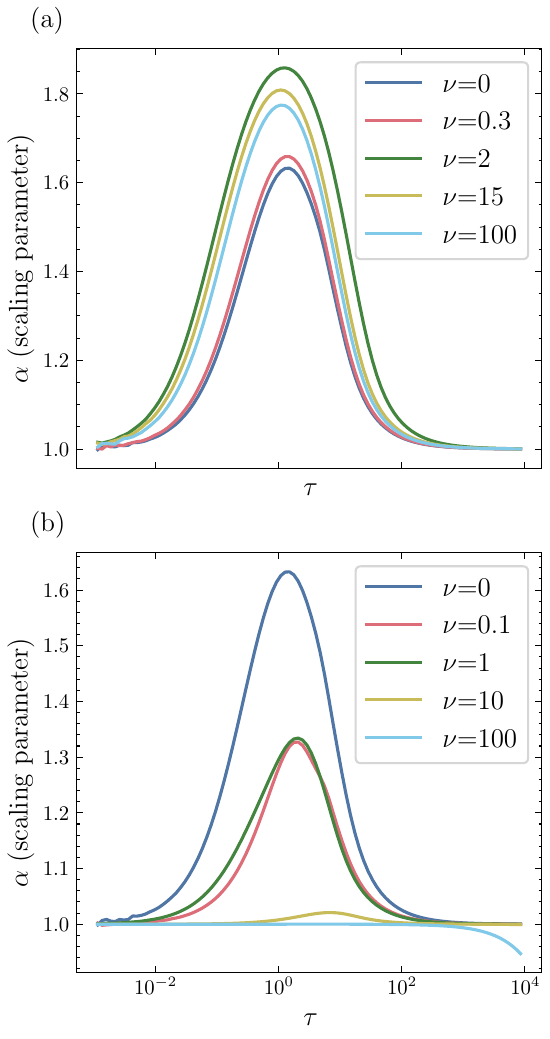}
\caption{\label{fig:qda_time_der} Numerical calculation of the time scaling order for quantum dynamical activity under feedback control. (a) and (b) show the results for feedback by jump and homodyne measurement, respectively. The different curves represent the results for various feedback strengths $ν$. The parameters are set to $\Delta=1.0$, $\Omega=1.0$, $\kappa=0.5$, and $\phi_z=\pi/2$. 
} 
\end{figure}

We next quantitatively investigate the time-scaling orders of $\mathcal{B}_\mathrm{jmp}^\mathrm{fb}(\tau)$ and $\mathcal{B}_\mathrm{hom}^\mathrm{fb}(\tau)$ to provide a more detailed discussion on the impact of feedback control. This scaling behavior is of fundamental importance because it dictates the time dependence of performance limits, such as those established by QSL and quantum TUR. To characterize this behavior precisely, we assume a power-law scaling form, $\mathcal{B}_\mathrm{jmp(hom)}^\mathrm{fb}(\tau)\propto \tau^\alpha$, where $\alpha$ is a real exponent. The scaling parameter $\alpha$ directly quantifies the rate of improvement; for example, $\alpha$ signifies that the precision limit improves proportionally to $\tau^\alpha$. We present plots of $\alpha$ for $\mathcal{B}_\mathrm{jmp}^\mathrm{fb}(\tau)$ and $\mathcal{B}_\mathrm{hom}^\mathrm{fb}(\tau)$ in \Fref{fig:qda_time_der}(a) and (b), respectively. We note that $\nu=0$ corresponds to the scenario without feedback. For feedback control by jump measurements [\Fref{fig:qda_time_der}(a)], it is evident that the presence of feedback consistently enhances the time scaling order. Furthermore, we observe a positive correlation between larger values of $\alpha$ and $\nu$ which yield a larger $\mathcal{B}_\mathrm{jmp}^\mathrm{fb}(\tau)$. In contrast, feedback control by homodyne measurement exhibits the different effect [\Fref{fig:qda_time_der}(b)], where the presence of feedback degrades the scaling behavior. Specifically, while increasing the feedback strength $\nu$ successfully increases $\mathcal{B}_\mathrm{hom}^\mathrm{fb}(\tau)$ itself, the corresponding time scaling progressively deteriorates.

\subsection{Feedback Strength Dependence}\label{strength_dependence}
\begin{figure}[t]
\includegraphics[width=70mm]{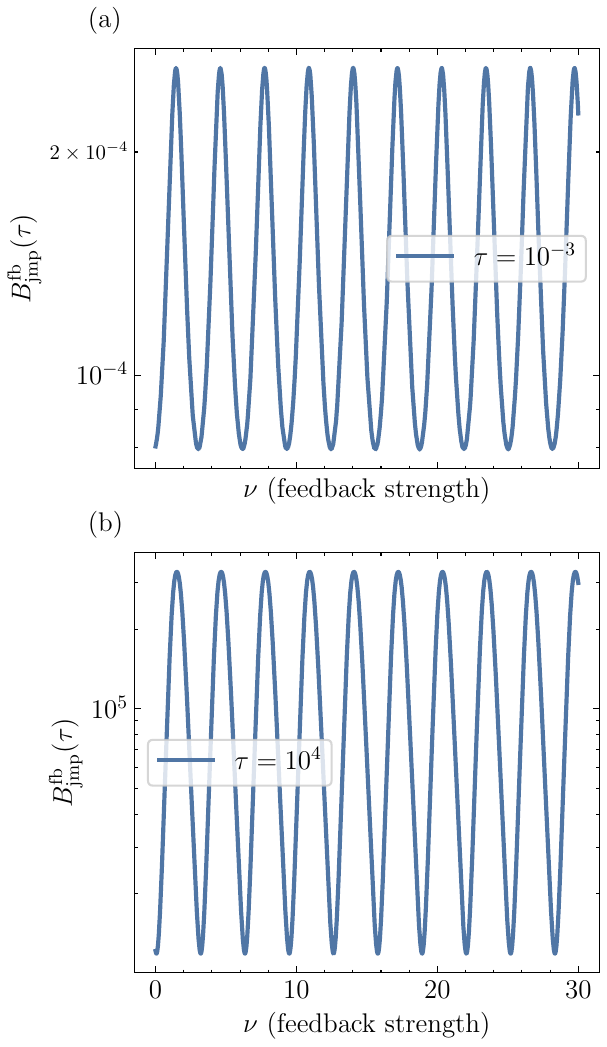}
\caption{\label{fig:qda_strength_jmp} Numerical calculation of the feedback strength dependence of quantum dynamical activity under feedback control by jump measurement. (a) and (b) show the cases of $\tau=10^{-3}$ and $\tau=10^4$, respectively. The parameters are set to $\Delta=1.0$, $\Omega=1.0$, $\kappa=0.5$. 
} 
\end{figure}

\begin{figure}[t]
\includegraphics[width=70mm]{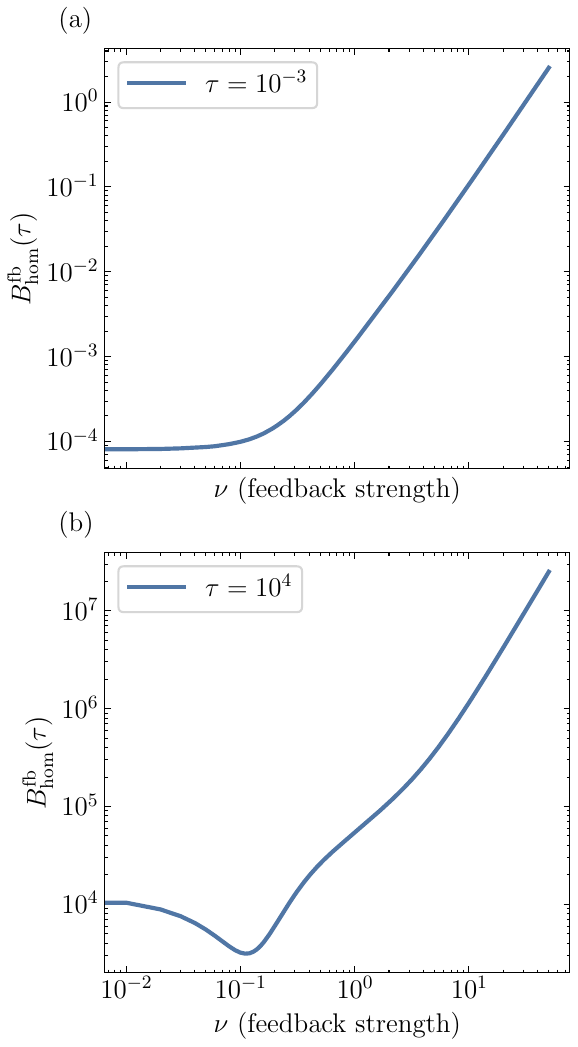}
\caption{\label{fig:qda_strength_hom} Numerical calculation of the feedback strength dependence of quantum dynamical activity under feedback control by homodyne measurement. (a) and (b) show the cases of $\tau=10^{-3}$ and $\tau=10^4$, respectively. The parameters are set to $\Delta=1.0$, $\Omega=1.0$, $\kappa=0.5$, and $\phi_z=\pi/2$. 
} 
\end{figure}

Next, we investigate the feedback strength $\nu$ dependence of the quantum dynamical activity under feedback control. Figures \ref{fig:qda_strength_jmp} and \ref{fig:qda_strength_hom} illustrate this dependence for feedback by jump and homodyne measurements, respectively. As in the previous analysis, the calculations are performed on the two-level atom driven by a classical laser field. In each figure, panels (a) and (b) show the behavior in the short-time ($\tau=10^{-3}$) and long-time ($\tau=10^4$) regimes, respectively. Note again that the case $\nu=0$ corresponds to the dynamics without feedback control.

For the jump measurement case shown in \Fref{fig:qda_strength_jmp}, the quantum dynamical activity $\mathcal{B}_\mathrm{jmp}^\mathrm{fb}(\tau)$ exhibits a clear oscillatory pattern as a function of the feedback strength $\nu$. Notably, the frequency of this oscillation is independent of $\tau$, and $\mathcal{B}_\mathrm{jmp}^\mathrm{fb}(\tau)$ is consistently minimized in the absence of feedback ($\nu=0$). The origin of this behavior can be traced back to our analytical NH-type solution of $\mathcal{B}_\mathrm{jmp}^\mathrm{fb}(\tau)$ [\Eref{Nishiyama_jump_QDA}]. The sole difference in this expression compared to the no-feedback case $\mathcal{B}(\tau)$ is the term $e^{\nu_z \mathcal{F}^\dagger}(\mathcal{O})=e^{i\nu_z F}\mathcal{O}e^{-i\nu_zF}$ within the adjoint superoperator $\mathcal{L}_\mathrm{J}^\dagger$. This term is responsible for the periodic behavior of $\mathcal{B}_\mathrm{jmp}^\mathrm{fb}(\tau)$ as a function of $\nu$, and its frequency is governed solely by the feedback operator $F$, thus explaining its independence from $\tau$. Furthermore, this explains the fact that was observed in the time dependence plot of \Fref{fig:qda_time}(a), indicating the non-monotonic behavior of $\mathcal{B}_\mathrm{jmp}^\mathrm{fb}(\tau)$ with respect to $\nu$.

Next, we focus on the homodyne measurement case in \Fref{fig:qda_strength_hom}. The behavior differs significantly depending on $\tau$. In the short-time regime [\Fref{fig:qda_strength_hom}(a)], $\mathcal{B}_\mathrm{hom}^\mathrm{fb}(\tau)$ is a monotonically increasing function of $\nu$. In the long-time regime [\Fref{fig:qda_strength_hom}(b)], however, it exhibits non-monotonic behavior, featuring a local minimum at a small, non-zero value of $\nu$. The existence of this local minimum implies that, in this regime, the quantum dynamical activity is not necessarily enhanced by the contribution of feedback control. This finding precisely explains our earlier observation from the time dependence plot [\Fref{fig:qda_time}(b)], where the general increase of activity with $\nu$ failed in the specific regime of large $\tau$ and small $\nu$.

\section{Conclusion}
In this study, we derived the fundamental trade-off relations that describe the theoretical performance limits in non-equilibrium quantum systems, QSL and quantum TUR, under feedback control by jump, homodyne, and Gaussian measurement. Additionally, we derived exact analytical expressions for the quantum dynamical activity, the thermodynamic cost constraining speed and precision, in terms of tangible physical quantities for each feedback scheme. Numerical simulations verified that our derived inequalities hold and demonstrated that the presence of feedback control can enhance the quantum dynamical activity, leading to the improved speed and precision. We further showcased the applicability of our results by investigating the quantum TUR in the context of quantum error correction, which is one of the most important applications of quantum feedback control. Here again, we confirmed the validity of the inequality and the enhancement of precision due to the feedback contribution. To provide a deeper understanding of the cost term itself, we also numerically investigated the dependence of our exact representation of quantum dynamical activity on time and feedback strength in a simple system. In particular, our analysis revealed that feedback control by jump measurement can improve the very order of the time scaling of quantum dynamical activity. This finding indicates that the presence of feedback can enhance the fundamental scaling of metrics related to speed and precision, not merely their numerical values. Our results clarify that active manipulation through quantum feedback control can improve the fundamental performance limits of quantum systems. Feedback control in quantum systems is a crucial technique from both fundamental and applied perspectives. Our study contributes to a deeper understanding of quantum dynamics under feedback control and may facilitate future developments in quantum technologies.

\section*{ACKNOWLEDGMENTS}
This work was supported by JSPS KAKENHI Grant No. JP23K24915

\appendix

\section{Liouville space representation}
An arbitrary linear operator $A$ in Hilbert space can be described as follows:
\begin{equation}
    A=\sum_{i, j}A_{ij}|i\rangle\langle j|,
\end{equation}
where $|i\rangle$ is the orthonormal basis in the Hilbert space. $A$ can become a vectorized form $|A\rangle\!\rangle$ defined by
\begin{equation}\label{liouville_def}
    |A\rangle\!\rangle\equiv\sum_{i, j} A_{ij}|j\rangle\otimes|i\rangle,
\end{equation}
which belongs to a Liouville space. From \Eref{liouville_def}, we can obtain the following relation:
\begin{equation}\label{liouville_relation}
    |ABC\rangle\!\rangle=(C^\mathsf{T}\otimes A)|B\rangle\!\rangle,
\end{equation}
where $\mathsf{T}$ means matrix transpose. The inner product of these vectors is described by
\begin{equation}
    \langle\!\langle B|A\rangle\!\rangle=\mathrm{Tr}[B^\dagger A].
\end{equation}
Specifically $\langle\!\langle 1|A\rangle\!\rangle=\mathrm{Tr}[A]$ holds. When $\mathcal{L}$ is the superoperator in the equation describing quantum dynamics, the following holds:
\begin{equation}\label{liouville_conserve}
    \langle\!\langle 1|\hat{\mathcal{L}}=0,
\end{equation}
from conservation of probability, where $\hat{\bullet}$ is the Liouville space representation of super operator $\bullet$. Using \Eref{liouville_def} and \Eref{liouville_relation}, in Liouville space, the equation describing quantum dynamics becomes
\begin{equation}
    \frac{d|\rho(t)\rangle\!\rangle}{dt}=\hat{\mathcal{L}}|\rho(t)\rangle\!\rangle.
\end{equation}
 When $\mathcal{L}$ is time independent, we can obtain
\begin{equation}\label{liouville_evolve}
    |\rho(t)\rangle\!\rangle=\exp{(\hat{\mathcal{L}}(t-s))}|\rho(s)\rangle\!\rangle.
\end{equation}
From \Eref{liouville_conserve}, we have
\begin{equation}\label{liouville_conserve2}
    \langle\!\langle1|\exp{(\hat{\mathcal{L}}t)}=\langle\!\langle1|.
\end{equation}

\begin{widetext}
\section{Derivation of Quantum Dynamical Activity under Feedback Control by Jump Measurement}\label{jump_derivation}
In this section, we provide the derivation of quantum dynamical activity under feedback control by jump measurement in two forms[\Eref{Nakajima_jump_QDA}, \Eref{Nishiyama_jump_QDA}] based on the methods in Ref. \cite{nakajimaSymmetriclogarithmicderivativeFisherInformation2023, nishiyamaExactSolutionQuantum2024}. 

At first, we derive NU-type of quantum dynamical activity [\Eref{Nakajima_jump_QDA}]. When we define $C(\theta, \phi)$ as
\begin{equation}\label{def_c}
    C(\theta, \phi)\equiv \mathrm{Tr}_S[\rho^{\theta, \phi}(\tau)],
\end{equation}
where $\rho^{\theta, \phi}(\tau)$ follows \Eref{jump_two_sided}, quantum dynamical activity $\mathcal{B}_{\mathrm{jmp}}^{\mathrm{fb}}(\tau)$ we want can be described as follows:
\begin{equation}\label{what_calc_jump}
    \mathcal{B}_{\mathrm{jmp}}^{\mathrm{fb}}(\tau)=4[\partial_\theta \partial_\phi C(\theta, \phi)-\partial_\theta C(\theta, \phi) \partial_\phi C(\theta, \phi)]|_{\theta=\phi=0}.
\end{equation}
From \Eref{liouville_evolve} and \Eref{def_c}, we obtain
\begin{equation}
    C(\theta, \phi)=\langle\!\langle1|\exp({\hat{\mathcal{L}}_\mathrm{J}^{\theta, \phi}}\tau)|\rho^{\theta, \phi}(0)\rangle\!\rangle.
\end{equation}
The first derivative of $C(\theta, \phi)$ becomes
\begin{equation}\label{first_der_j}
    \partial_{\theta_i}C(\theta, \phi)=\int_0^\tau du\langle\!\langle1|\exp(\hat{\mathcal{L}}_\mathrm{J}^{\theta, \phi}(\tau-u))
    \partial_{\theta_i}\hat{\mathcal{L}}_\mathrm{J}^{\theta, \phi}\exp(\hat{\mathcal{L}}_\mathrm{J}^{\theta, \phi}u)|\rho^{\theta, \phi}(0)\rangle\!\rangle,
\end{equation}
where $\theta_i=\theta, \phi$. From \Eref{liouville_conserve2}, $\partial_{\theta_i}C(\theta, \phi)|_{\theta=\phi=0}$ can be written by
\begin{equation}
    \partial_{\theta_i}C(\theta, \phi)|_{\theta=\phi=0}=\int_0^\tau ds \langle\!\langle1|\partial_{\theta_i} \hat{\mathcal{L}}_\mathrm{J}^{\theta, \phi}|\rho(s)\rangle\!\rangle|_{\theta=\phi=0}.
\end{equation}
By calculating this, we can obtain
\begin{equation}
    \partial_{\theta}C(\theta, \phi)|_{\theta=\phi=0}=\int_0^\tau ds \mathrm{Tr}_S[\mathcal{K}_{\mathrm{J}1}\rho(s)],
\end{equation}
\begin{equation}
    \partial_{\phi}C(\theta, \phi)|_{\theta=\phi=0}=\int_0^\tau ds \mathrm{Tr}_S[\mathcal{K}_{\mathrm{J}2}\rho(s)],
\end{equation}
where $\mathcal{K}_{\mathrm{J}1}$ and $\mathcal{K}_{\mathrm{J}2}$ are defined in \Eref{k_j1} and \Eref{k_j2}. Then we can calculate the second term of \Eref{what_calc_jump} as follows:
\begin{equation}\label{jump_qda_2}
\begin{split}
    -\partial_\theta C(\theta, \phi)\partial_\phi C(\theta, \phi)|_{\theta=\phi=0}&=-\Pi_{i=1}^2 \int_0^\tau ds \mathrm{Tr}[\mathcal{K}_{\mathrm{J}i}\rho(s)]\\
    &=-\left\{\int_0^\tau ds \big(-i\mathrm{Tr}_S[H\rho(s)]\big)+\frac{1}{2}\sum_z \big(\mathrm{Tr}_S[e^{\nu_z\mathcal{F}}(L_z\rho(s)L_z^\dagger)]-\mathrm{Tr}_S[L_z^\dagger L_z \rho(s)] \big)\right\}^2\\
    &= -\left(\int_0^\tau ds \mathrm{Tr}_S[H\rho(s)]\right)^2.
\end{split}
\end{equation}
From \Eref{first_der_j}, the first term of \Eref{what_calc_jump} becomes
\begin{equation}\label{second_der_j}
\begin{split}
    \partial_{\theta}\partial_{\phi}C(\theta, \phi)|_{\theta=\phi=0}
    =&\int_0^\tau du \langle\!\langle1|\partial_\theta\exp(\hat{\mathcal{L}}_\mathrm{J}^{\theta, \phi}(\tau-u))\partial_{\phi}\hat{\mathcal{L}}_\mathrm{J}^{\theta, \phi}\exp(\hat{\mathcal{L}}_\mathrm{J}^{\theta, \phi}u)\\
    &+\exp(\hat{\mathcal{L}}_\mathrm{J}^{\theta, \phi}(\tau-u))\partial_\theta\partial_{\phi}\hat{\mathcal{L}}_\mathrm{J}^{\theta, \phi}\exp(\hat{\mathcal{L}}_\mathrm{J}^{\theta, \phi}u)\\
    &+\exp(\hat{\mathcal{L}}_\mathrm{J}^{\theta, \phi}(\tau-u))\partial_{\phi}\hat{\mathcal{L}}_\mathrm{J}^{\theta, \phi}\partial_\theta\exp(\hat{\mathcal{L}}_\mathrm{J}^{\theta, \phi}u)|\rho^{\theta, \phi}(0)\rangle\!\rangle|_{\theta=\phi=0}.
\end{split}
\end{equation}
The second term of \Eref{second_der_j} can be calculated as follows
\begin{equation}\label{second_term_j}
    \int_0^\tau du\langle\!\langle1|\partial_\theta\partial_{\phi}\hat{\mathcal{L}}_\mathrm{J}^{\theta, \phi}|\rho^{\theta, \phi}(s)\rangle\!\rangle|_{\theta=\phi=0}=\frac{1}{4}\mathcal{A}(\tau).
\end{equation}
The first term of \Eref{second_der_j} becomes
\begin{equation}\label{first_term_j}
    \int_0^\tau du \langle\!\langle1|\int_u^\tau ds \exp(\hat{\mathcal{L}}_\mathrm{J}^{\theta, \phi}(\tau-s))\partial_{\theta}\hat{\mathcal{L}}_\mathrm{J}^{\theta, \phi}
    \exp(\hat{\mathcal{L}}_\mathrm{J}^{\theta, \phi}(s-u))\partial_{\phi}\hat{\mathcal{L}}_\mathrm{J}^{\theta, \phi}\exp(\hat{\mathcal{L}}_\mathrm{J}^{\theta, \phi}(u))|\rho^{\theta, \phi}(0)\rangle\!\rangle|_{\theta=\phi=0}
    =I_{\mathrm{J}2}.
\end{equation}
The third term of \Eref{second_der_j} becomes
\begin{equation}\label{third_term_j}
    \int_0^\tau du \langle\!\langle1|\exp(\hat{\mathcal{L}}_\mathrm{J}^{\theta, \phi}(\tau-u))\partial_{\phi}\hat{\mathcal{L}}_\mathrm{J}^{\theta, \phi}
  \int_0^u ds \exp(\hat{\mathcal{L}}_\mathrm{J}^{\theta, \phi}(s-u))\partial_{\theta}\hat{\mathcal{L}}_\mathrm{J}^{\theta, \phi}\exp(\hat{\mathcal{L}}_\mathrm{J}^{\theta, \phi}(u))|\rho^{\theta, \phi}(0)\rangle\!\rangle|_{\theta=\phi=0}
    =I_{\mathrm{J}1}.
\end{equation}
Then, we obtain the first term of \Eref{what_calc_jump} as follows:
\begin{equation}\label{jump_qda_1}
    \partial_{\theta}\partial_{\phi}C(\theta, \phi)|_{\theta=\phi=0}=\frac{1}{4}\mathcal{A}(\tau)+I_{\mathrm{J}1}+I_{\mathrm{J}2}.
\end{equation}
From \Eref{jump_qda_1} and \Eref{jump_qda_2}, we obtain NU-type of quantum dynamical activity [\Eref{Nakajima_jump_QDA}].

Next, we can derive NH-type of quantum dynamical activity [\Eref{Nishiyama_jump_QDA}] from NU-type of quantum dynamical activity. From the cyclic property of trace, we can obtain the following relations:
\begin{equation}
    \mathrm{Tr}_S[\mathcal{K}_{\mathrm{J}1}\bullet]=-i\mathrm{Tr}_S[H\bullet],
\end{equation}
\begin{equation}
    \mathrm{Tr}_S[\mathcal{K}_{\mathrm{J}2}\bullet]=i\mathrm{Tr}_S[H\bullet].
\end{equation}
By applying these relations, $I_{\mathrm{J}1}$ and $I_{\mathrm{J}2}$ become
\begin{equation}
    I_{\mathrm{J}1}=i\int_0^\tau ds \int_0^s du \mathrm{Tr}_S[H\exp(\mathcal{L}_\mathrm{J}(s-u))\mathcal{K}_{\mathrm{J}1}\rho(u)]
\end{equation}
\begin{equation}
    I_{\mathrm{J}2}=-i\int_0^\tau ds \int_0^s du \mathrm{Tr}_S[H\exp(\mathcal{L}_\mathrm{J}(s-u))\mathcal{K}_{\mathrm{J}2}\rho(u)].
\end{equation}
By representing $\exp(\mathcal{L}_\mathrm{J}(s-u))$ with Kraus operators $M_z$ and $U_z$ and using cyclic property of trace, we have
\begin{equation}
    I_{\mathrm{J}1}=\int_0^\tau ds \int_0^{s} du \operatorname{Tr}_S[\breve{H}(s - u)H_{\text{eff}} \rho(u)]
+ \frac{i}{2} \int_0^\tau ds \int_0^{s} du \sum_k \operatorname{Tr}_S[\breve{H}(s - u)e^{\nu_z\mathcal{F}}L_k\rho(u)L_k^\dagger],
\end{equation}
\begin{equation}
    I_{\mathrm{J}2}=\int_0^\tau ds \int_0^{s} du \operatorname{Tr}_S[\breve{H}(s - u) \rho(u)H_{\text{eff}}^\dagger]
- \frac{i}{2} \int_0^\tau ds \int_0^{s} du \sum_k \operatorname{Tr}_S[\breve{H}(s - u)e^{\nu_z\mathcal{F}}L_k\rho(u)L_k^\dagger],
\end{equation}
where $\breve{\bullet}$ is given by
\begin{equation}
    \breve{\bullet}=\sum_{\bm{z}} M_{z_0}^\dagger U_{z_0}^\dagger\cdots M_{z_{N_\tau-1}}^\dagger U_{z_{N_\tau-1}}^\dagger\bullet U_{z_{N_\tau-1}}M_{z_{N_\tau-1}}\cdots U_{z_0}M_{z_0},
\end{equation}
when time interval $[u, s]$ is divided by large $N_\tau$. Given that $\bullet(t)$ evolves by Kraus operator $U_z^\dagger$ and $M_z^\dagger$ during infinitesimal time to obtain the concrete form of $\breve{\bullet}$, we obtain
\begin{equation}
\begin{split}
    \bullet(t+dt)&=(1-\frac{1}{2}\sum_z L_z^\dagger L_zdt)e^{iHdt}\bullet(t)e^{-iHdt}(1-\frac{1}{2}\sum_z L_z^\dagger L_zdt)+\sum_z \sqrt{dt}L_z^\dagger e^{i\nu_z F}e^{iHdt}\bullet e^{-iHdt}e^{-i\nu_z F}\sqrt{dt}L_z\\
    &=\bullet(t)+i[H, \bullet(t)]+\sum_zL_z^\dagger e^{i\nu_z F}\bullet(t)e^{-i\nu_z F}L_z-\frac{1}{2}\bullet(t)L_z^\dagger L_z-\frac{1}{2}L_z^\dagger L_z \bullet(t)+O(dt^2).
\end{split}
\end{equation}
Then we can derive the following equation:
\begin{equation}\label{jump_dag_evolve}
    \dot{\bullet}= i[H, \bullet]+\sum_{z=1}^{N_c}L_z^\dag e^{\nu_z \mathcal{F}^\dagger}(\bullet)L_z-\frac{1}{2}\{L_z^\dag L_z, \bullet\}\equiv\mathcal{L}_{\mathrm{J}}^\dag\bullet.
\end{equation}
From \Eref{jump_dag_evolve}, we can get the definition of $\breve{\bullet}$ as follows:
\begin{equation}
    \breve{\bullet}(t)\equiv \exp(\mathcal{L}_{\mathrm{J}}^\dagger t)\bullet.
\end{equation}
By adding $I_{\mathrm{J}1}$ and $I_{\mathrm{J}2}$, we can obtain
\begin{equation}
    I_{\mathrm{J}1} + I_{\mathrm{J}2} = 2 \int_0^\tau ds \int_0^{s} du \operatorname{Re}\left(\operatorname{Tr}_S\left[H_{\text{eff}}^\dagger \breve{H}(s - u) \rho(u)\right]\right)
\end{equation}
Then, we can get NH-type of quantum dynamical activity [\Eref{Nishiyama_jump_QDA}].

\section{Derivation of Quantum Dynamical Activity under Feedback Control by Homodyne Measurement}\label{homodyne_derivation}
In this section, we provide the derivation of quantum dynamical activity under feedback control by homodyne measurement in two forms[\Eref{Nakajima_homodyne_QDA}, \Eref{Nishiyama_homodyne_QDA}] based on the methods in Ref. \cite{nakajimaSymmetriclogarithmicderivativeFisherInformation2023, nishiyamaExactSolutionQuantum2024}. 

At first, we derive NU-type of quantum dynamical activity [\Eref{Nakajima_homodyne_QDA}]. When we define $C(\theta, \phi)$ as
\begin{equation}\label{def_homodyne_c}
    C(\theta, \phi)\equiv \mathrm{Tr}_S[\rho^{\theta, \phi}(\tau)],
\end{equation}
where $\rho^{\theta, \phi}(\tau)$ follows \Eref{homodyne_two_sided}, quantum dynamical activity $\mathcal{B}_{\mathrm{hom}}^{\mathrm{fb}}(\tau)$ we want can be described as follows:
\begin{equation}\label{what_calc_homodyne}
    \mathcal{B}_{\mathrm{hom}}^{\mathrm{fb}}(\tau)=4[\partial_\theta \partial_\phi C(\theta, \phi)-\partial_\theta C(\theta, \phi) \partial_\phi C(\theta, \phi)]|_{\theta=\phi=0}.
\end{equation}
From \Eref{liouville_evolve} and \Eref{def_homodyne_c}, we obtain
\begin{equation}
    C(\theta, \phi)=\langle\!\langle1|\exp({\hat{\mathcal{L}}_\mathrm{H}^{\theta, \phi}}\tau)|\rho^{\theta, \phi}(0)\rangle\!\rangle.
\end{equation}
The first derivative of $C(\theta, \phi)$ becomes
\begin{equation}\label{first_der_h}
    \partial_{\theta_i}C(\theta, \phi)=\int_0^\tau du\langle\!\langle1|\exp(\hat{\mathcal{L}}_\mathrm{H}^{\theta, \phi}(\tau-u))\partial_{\theta_i}\hat{\mathcal{L}}_\mathrm{H}^{\theta, \phi}\exp(\hat{\mathcal{L}}_\mathrm{H}^{\theta, \phi}u)|\rho^{\theta, \phi}(0)\rangle\!\rangle,
\end{equation}
where $\theta_i=\theta, \phi$. From \Eref{liouville_conserve2}, $\partial_{\theta_i}C(\theta, \phi)|_{\theta=\phi=0}$ can be written by
\begin{equation}
    \partial_{\theta_i}C(\theta, \phi)|_{\theta=\phi=0}=\int_0^\tau ds \langle\!\langle1|\partial_{\theta_i} \hat{\mathcal{L}}_\mathrm{H}^{\theta, \phi}|\rho(s)\rangle\!\rangle|_{\theta=\phi=0}.
\end{equation}
By calculating this, we can obtain
\begin{equation}
    \partial_{\theta}C(\theta, \phi)|_{\theta=\phi=0}=\int_0^\tau ds \mathrm{Tr}_S[\mathcal{K}_{\mathrm{H}1}\rho(s)],
\end{equation}
\begin{equation}
    \partial_{\phi}C(\theta, \phi)|_{\theta=\phi=0}=\int_0^\tau ds \mathrm{Tr}_S[\mathcal{K}_{\mathrm{H}2}\rho(s)],
\end{equation}
where $\mathcal{K}_{\mathrm{H}1}$ and $\mathcal{K}_{\mathrm{H}2}$ are defined in \Eref{k_h1} and \Eref{k_h2}. Then we can calculate the second term of \Eref{what_calc_homodyne} as follows:
\begin{equation}\label{homodyne_qda_2}
\begin{split}
    -&\partial_\theta C(\theta, \phi)\partial_\phi C(\theta, \phi)|_{\theta=\phi=0}\\
    =&-\Pi_{i=1}^2 \int_0^\tau ds \mathrm{Tr}[\mathcal{K}_{\mathrm{H}i}\rho(s)]\\
    =&\left\{-i\int_0^\tau ds \big(\mathrm{Tr}_S[H\rho(s)]+\sum_z\frac{1}{2}\mathrm{Tr}_S[e^{i\phi_z}F\rho(s)L_z^\dagger+e^{-i\phi_z}FL_z\rho(s)]\big)\right\}\\
    &\times\left\{i\int_0^\tau ds \big(\mathrm{Tr}_S[H\rho(s)]+\sum_z\frac{1}{2}\mathrm{Tr}_S[e^{i\phi_z}F\rho(s)L_z^\dagger+e^{-i\phi_z}FL_z\rho(s)]\big)\right\}\\
    =& -\left\{\int_0^\tau ds \big(\mathrm{Tr}_S[H\rho(s)]+\sum_z\frac{1}{2}\mathrm{Tr}_S[e^{i\phi_z}F\rho(s)L_z^\dagger+e^{-i\phi_z}FL_z\rho(s)]\big)\right\}^2.
\end{split}
\end{equation}
From \Eref{first_der_h}, the first term of \Eref{what_calc_homodyne} becomes
\begin{equation}
\begin{split}\label{second_der_h}
    \partial_{\theta}\partial_{\phi}C(\theta, \phi)|_{\theta=\phi=0}=&\int_0^\tau du \langle\!\langle1|\partial_\theta\exp(\hat{\mathcal{L}}_\mathrm{H}^{\theta, \phi}(\tau-u))\partial_{\phi}\hat{\mathcal{L}}_\mathrm{H}^{\theta, \phi}\exp(\hat{\mathcal{L}}_\mathrm{H}^{\theta, \phi}u)\\
    &+\exp(\hat{\mathcal{L}}_\mathrm{H}^{\theta, \phi}(\tau-u))\partial_\theta\partial_{\phi}\hat{\mathcal{L}}_\mathrm{H}^{\theta, \phi}\exp(\hat{\mathcal{L}}_\mathrm{H}^{\theta, \phi}u)\\
    &+\exp(\hat{\mathcal{L}}_\mathrm{H}^{\theta, \phi}(\tau-u))\partial_{\phi}\hat{\mathcal{L}}_\mathrm{H}^{\theta, \phi}\partial_\theta\exp(\hat{\mathcal{L}}_\mathrm{H}^{\theta, \phi}u)|\rho^{\theta, \phi}(0)\rangle\!\rangle|_{\theta=\phi=0}.
\end{split}
\end{equation}
The second term of \Eref{second_der_h} can be calculated as follows
\begin{equation}\label{second_term_h}
    \int_0^\tau du\langle\!\langle1|\partial_\theta\partial_{\phi}\hat{\mathcal{L}}_\mathrm{H}^{\theta, \phi}|\rho^{\theta, \phi}(s)\rangle\!\rangle|_{\theta=\phi=0}=\frac{1}{4}\mathcal{A}_{\mathrm{hom}}^{\mathrm{fb}}(\tau).
\end{equation}
The first term of \Eref{second_der_h} becomes
\begin{equation}\label{first_term_h}
    \int_0^\tau du \langle\!\langle1|\int_u^\tau ds \exp(\hat{\mathcal{L}}_\mathrm{H}^{\theta, \phi}(\tau-s))\partial_{\theta}\hat{\mathcal{L}}_\mathrm{H}^{\theta, \phi}
    \exp(\hat{\mathcal{L}}_\mathrm{H}^{\theta, \phi}(s-u))\partial_{\phi}\hat{\mathcal{L}}_\mathrm{H}^{\theta, \phi}\exp(\hat{\mathcal{L}}_\mathrm{H}^{\theta, \phi}(u))|\rho^{\theta, \phi}(0)\rangle\!\rangle|_{\theta=\phi=0}
    =I_{\mathrm{H}2}.
\end{equation}
The third term of \Eref{second_der_h} becomes
\begin{equation}\label{third_term_h}
    \int_0^\tau du \langle\!\langle1|\exp(\hat{\mathcal{L}}_\mathrm{H}^{\theta, \phi}(\tau-u))\partial_{\phi}\hat{\mathcal{L}}_\mathrm{H}^{\theta, \phi}
    \int_0^u ds \exp(\hat{\mathcal{L}}_\mathrm{H}^{\theta, \phi}(s-u))\partial_{\theta}\hat{\mathcal{L}}_\mathrm{H}^{\theta, \phi}\exp(\hat{\mathcal{L}}_\mathrm{H}^{\theta, \phi}(u))|\rho^{\theta, \phi}(0)\rangle\!\rangle|_{\theta=\phi=0}
    =I_{\mathrm{H}1}.
\end{equation}
Then, we obtain the first term of \Eref{what_calc_homodyne} as follows:
\begin{equation}\label{homodyne_qda_1}
    \partial_{\theta}\partial_{\phi}C(\theta, \phi)|_{\theta=\phi=0}=\frac{1}{4}\mathcal{A}_{\mathrm{hom}}^{\mathrm{fb}}(\tau)+I_{\mathrm{H}1}+I_{\mathrm{H}2}.
\end{equation}
From \Eref{homodyne_qda_1} and \Eref{homodyne_qda_2}, we obtain NU-type of quantum dynamical activity [\Eref{Nakajima_homodyne_QDA}].

Next, we can derive NH-type of quantum dynamical activity [\Eref{Nishiyama_homodyne_QDA}] from NU-type of quantum dynamical activity. From the cyclic property of trace, we can obtain the following relations:
\begin{equation}
    \mathrm{Tr}_S[\mathcal{K}_{\mathrm{H}1}\bullet]=-i(\mathrm{Tr}_S[H\bullet]+\sum_z\frac{1}{2}\mathrm{Tr}_S[e^{i\phi_z}F\bullet L_z^\dagger+e^{-i\phi_z}FL_z\bullet]),
\end{equation}
\begin{equation}
    \mathrm{Tr}_S[\mathcal{K}_{\mathrm{H}1}\bullet]=i(\mathrm{Tr}_S[H\bullet]+\sum_z\frac{1}{2}\mathrm{Tr}_S[e^{i\phi_z}F\bullet L_z^\dagger+e^{-i\phi_z}FL_z\bullet]).
\end{equation}
By applying these relations, $I_{\mathrm{H}1}$ and $I_{\mathrm{H}2}$ become
\begin{equation}
\begin{split}
    I_{\mathrm{H}1}=&i\int_0^\tau ds \int_0^s du \Big\{\mathrm{Tr}_S[H\exp(\mathcal{L}_\mathrm{H}(s-u))\mathcal{K}_{\mathrm{H}1}\rho(u)]\\
    &+\sum_z\frac{1}{2}\mathrm{Tr}[e^{i\phi_z}F\exp(\mathcal{L}_{\mathrm{H}}(s-u))(\mathcal{K}_{\mathrm{H}1}\rho(u))L_z^\dagger+e^{-i\phi_z}FL_z\exp(\mathcal{L}_{\mathrm{H}}(s-u))(\mathcal{K}_{\mathrm{H}1}\rho(u))]\Big\}
\end{split}
\end{equation}
\begin{equation}
\begin{split}
    I_{\mathrm{H}2}=&-i\int_0^\tau ds \int_0^s du \Big\{\mathrm{Tr}_S[H\exp(\mathcal{L}_\mathrm{H}(s-u))\mathcal{K}_{\mathrm{H}2}\rho(u)]\\
    &+\sum_z\frac{1}{2}\mathrm{Tr}[e^{i\phi_z}F\exp(\mathcal{L}_{\mathrm{H}}(s-u))(\mathcal{K}_{\mathrm{H}2}\rho(u))L_z^\dagger+e^{-i\phi_z}FL_z\exp(\mathcal{L}_{\mathrm{H}}(s-u))(\mathcal{K}_{\mathrm{H}2}\rho(u))]\Big\}.
\end{split}
\end{equation}
By representing $\exp(\mathcal{L}_\mathrm{H}(s-u))$ with Kraus operators $M_z$ and $U_z$ and using cyclic property of trace, we have
\begin{equation}
\begin{split}
    I_{\mathrm{H}1}=&i\int_0^\tau ds \int_0^{s} du \Big\{\operatorname{Tr}_S[\widetilde{H}(s - u)\mathcal{K}_{\mathrm{H1}} \rho(u)]\\
    &+ \sum_z\frac{1}{2}\mathrm{Tr}_S[e^{i\phi_z}\widetilde{L_z^\dagger F}(s-u)(\mathcal{K}_{\mathrm{H}1}\rho(u))]+ \frac{1}{2}\mathrm{Tr}_S[e^{-i\phi_z}\widetilde{FL_z}(s-u)(\mathcal{K}_{\mathrm{H}1}\rho(u))] \Big\},
\end{split}
\end{equation}
\begin{equation}
\begin{split}
    I_{\mathrm{H}2}=&-i\int_0^\tau ds \int_0^{s} du \Big\{\operatorname{Tr}_S[\widetilde{H}(s - u)\mathcal{K}_{\mathrm{H2}} \rho(u)]\\
    &+ \sum_z\frac{1}{2}\mathrm{Tr}_S[e^{i\phi_z}\widetilde{L_z^\dagger F}(s-u)(\mathcal{K}_{\mathrm{H}2}\rho(u))]+ \frac{1}{2}\mathrm{Tr}_S[e^{-i\phi_z}\widetilde{FL_z}(s-u)(\mathcal{K}_{\mathrm{H}2}\rho(u))] \Big\},
\end{split}
\end{equation}
where $\widetilde{\bullet}$ is given by
\begin{equation}
    \widetilde{\bullet}=\sum_{\bm{z}} M_{z_0}^\dagger U_{z_0}^\dagger\cdots M_{z_{N_\tau-1}}^\dagger U_{z_{N_\tau-1}}^\dagger\bullet U_{z_{N_\tau-1}}M_{z_{N_\tau-1}}\cdots U_{z_0}M_{z_0},
\end{equation}
when time interval $[u, s]$ is divided by large $N_\tau$. Given that $\bullet(t)$ evolves by Kraus operator $U_z^\dagger$ and $M_z^\dagger$ during infinitesimal time to obtain the concrete form of $\widetilde{\bullet}$, we obtain
\begin{equation}
\begin{split}
    \bullet(t+dt)=&\sum_z M_z^\dagger e^{I_\mathrm{hom}(t)\mathcal{F}^\dagger dt}e^{\{\mathcal{H}^\dagger-\frac{1}{2}\sum_z(\alpha_z^* L_z-\alpha_z L_z^\dagger)\}dt}\bullet(t) M_z\\
    =&\sum_z M_z^\dagger e^{\sum_z\langle e^{-i\phi_zL_z+e^{i\phi_z}L_z} \rangle \mathcal{F}^\dagger dt} e^{\Delta W_z \mathcal{F}^\dagger}e^{\{\mathcal{H}^\dagger-\frac{1}{2}\sum_z(\alpha_z^* L_z-\alpha_z L_z^\dagger)\}dt}\bullet(t) M_z\\
    =&\sum_z  M_z^\dagger (1+\mathcal{H}^\dagger dt+\sum_z\langle e^{-i\phi_zL_z+e^{i\phi_z}L_z} \rangle\mathcal{F}^\dagger dt+\Delta W_z\mathcal{F}^\dagger+\frac{dt}{2}(\mathcal{F}^\dagger)^2)\\
    &\times(1-\frac{1}{2}\sum_z(\alpha_zL_z^\dagger-\alpha_z^*L_z)dt)\bullet(t)(1-\frac{1}{2}\sum_z(\alpha_z^*L_z-\alpha_zL_z^\dagger)dt) M_z+o(dt),
\end{split}
\end{equation}
where $dW$ is replaced by $\Delta W$ for clarity of the equation. When we convert variable $\Delta W$ to $I_\mathrm{hom}(t)$ by using \Eref{homodyne_current}, the following relation holds:
\begin{equation}
    \Delta W_z = I_\mathrm{hom}(t)dt-\sum_z\langle e^{-i\phi_zL_z+e^{i\phi_z}L_z} \rangle dt.
\end{equation}
Thus, we have
\begin{equation}
\begin{split}
    &\bullet(t+dt)\\
    &=\sum_z M_z^\dagger(1+\mathcal{H}^\dagger +I_\mathrm{hom}(t)\mathcal{F}^\dagger dt+\frac{(\mathcal{F})^2}{2}dt)(1-\frac{1}{2}\sum_z(\alpha_zL_z^\dagger-\alpha_z^*L_z)dt)\bullet(t)(1-\frac{1}{2}\sum_z(\alpha_z^*L_z-\alpha_zL_z^\dagger)dt) M_z+o(dt).
\end{split}
\end{equation}
First, we calculate the following part:
\begin{equation}
\begin{split}\label{homodyne_first_part}
    \sum_z M_z^\dagger(I_\mathrm{hom}(t)\mathcal{F}^\dagger dt)(1-\frac{1}{2}\sum_z(\alpha_zL_z^\dagger-\alpha_z^*L_z)dt)\bullet(t)(1-\frac{1}{2}\sum_z(\alpha_z^*L_z-\alpha_zL_z^\dagger)dt) M_z+o(dt).
\end{split}
\end{equation}
By using the following relations:
\begin{equation}
    I_\mathrm{hom}(t)=\frac{1}{|\alpha_z|}\left(\frac{dN_z}{dt}-|\alpha_z|^2\right), 
\end{equation}
\begin{equation}
    |\alpha_z|dt \ll 1,
\end{equation}
we can calculate \Eref{homodyne_first_part} as follows:
\begin{equation}
\begin{split}
    &\sum_z M_z^\dagger(I_\mathrm{hom}(t)\mathcal{F}^\dagger dt)(1-\frac{1}{2}\sum_z(\alpha_zL_z^\dagger-\alpha_z^*L_z)dt)\bullet(t)(1-\frac{1}{2}\sum_z(\alpha_z^*L_z-\alpha_zL_z^\dagger)dt) M_z+o(dt)\\
    &=\sum_z M_z^\dagger\frac{1}{|\alpha_z|}\left(\frac{dN_z}{dt}-|\alpha_z|^2\right)dt(\mathcal{F}^\dagger\bullet(t))M_z+o(dt)\\
    &=\sum_z (L_z^\dagger+\alpha_z^*)\frac{1}{|\alpha_z|}(\mathcal{F}^\dagger\bullet(t))(L_z+\alpha_z)dt+o(dt)\\
    &=\sum_z L_z^\dagger(\mathcal{F}^\dagger\bullet(t))e^{i\phi_z}+e^{-i\phi_z}(\mathcal{F}^\dagger\bullet(t))L_z+o(dt).
\end{split}
\end{equation}
Next, we calculate the following part:
\begin{equation}
\begin{split}
    &\sum_z M_z^\dagger(1+\mathcal{H}^\dagger +\frac{(\mathcal{F})^2}{2}dt)(1-\frac{1}{2}\sum_z(\alpha_zL_z^\dagger-\alpha_z^*L_z)dt)\bullet(t)(1-\frac{1}{2}\sum_z(\alpha_z^*L_z-\alpha_zL_z^\dagger)dt) M_z+o(dt)\\
    &=\sum_z M_z^\dagger(\bullet(t)-\frac{1}{2}\sum_z(\alpha_zL_z^\dagger\bullet(t)-\alpha_z^* L_z\bullet(t)+\alpha_z^*\bullet(t)L_z-\alpha_z\bullet(t)L_z^\dagger)dt+\mathcal{H}^\dagger\bullet(t)dt+\frac{(\mathcal{F}^\dagger)^2}{2}\bullet(t)dt)M_z+o(dt)\\
    &=\bullet(t)+\mathcal{H}^\dagger\bullet(t)dt+\sum_z L_z^\dagger\bullet(t)L_zdt-\frac{1}{2}\{L_z^\dagger L_z, \bullet(t)\}+\frac{(\mathcal{F})^2}{2}\bullet(t)dt+o(dt).
\end{split}
\end{equation}
Thus, we obtain
\begin{equation}
\begin{split}
    \bullet(t+dt)=&\bullet(t)+\mathcal{H}^\dagger\bullet(t) dt\\
    &+\sum_z L_z^\dagger\bullet(t) L_z dt-\frac{1}{2}\{L_z^\dagger L_z, \bullet(t)\}+L_z^\dagger(\mathcal{F}^\dagger\bullet(t))e^{i\phi_z}+e^{-i\phi_z}(\mathcal{F}^\dagger\bullet(t))L_z+\frac{(\mathcal{F})^2}{2}\bullet(t)dt+o(dt).
\end{split}
\end{equation}
Then we can derive the following equation:
\begin{equation}\label{homodyne_dag_evolve}
    \dot{\bullet}= i[H, \bullet]+\sum_zL_z^\dagger\bullet L_z-\frac{1}{2}\{L_z^\dagger L_z, \bullet\}+(e^{-i\phi_z}\mathcal{F}^\dagger (\bullet) L_z+e^{i\phi_z}L_z^\dagger\mathcal{F}^\dagger (\bullet))+\frac{(\mathcal{F}^\dagger)^2}{2}\bullet \equiv\mathcal{L}_{\mathrm{H}}^\dagger\bullet.
\end{equation}
From \Eref{homodyne_dag_evolve}, we can get the definition of $\widetilde{\bullet}$ as follows:
\begin{equation}
    \widetilde{\bullet}(t)\equiv \exp(\mathcal{L}_{\mathrm{H}}^\dagger t)\bullet.
\end{equation}
By using the following relation:
\begin{equation}
    i\mathcal{K}_{\mathrm{H}1}\bullet-i\mathcal{K}_{\mathrm{H}2}\bullet=H\bullet+\bullet H-\sum_z\frac{i}{2}L_z^\dagger L_z\bullet+\frac{i}{2}\bullet L_z^\dagger L_z+e^{-i\phi_z}FL_z\bullet+e^{i\phi_z}\bullet L_z^\dagger F-\frac{i}{2}F^2\bullet+\frac{i}{2}\bullet F,
\end{equation}
and $(e^{i\phi_z}\widetilde{L_z^\dagger F}(s-u))^\dagger=e^{-i\phi_z}\widetilde{FL_z}(s-u)$, we can obtain $I_{\mathrm{H}1} + I_{\mathrm{H}2}$ as follows:
\begin{equation}
    I_{\mathrm{H}1} + I_{\mathrm{H}2} = \int_0^\tau ds \int_0^{s} du \mathrm{Re}\left(2\operatorname{Tr}_S\left[\widetilde{H}(s - u)\rho(u)H_{\text{eff}}^{\mathrm{h}\dagger}+\sum_z\mathrm{Tr}_S[\{e^{i\phi_z}\widetilde{L_z^\dagger F}(s-u)+e^{-i\phi_z}\widetilde{FL_z}(s-u)\}\rho(u)H_{\text{eff}}^{\mathrm{h}\dagger}]  \right]\right).
\end{equation}
Then, we can get NH-type of quantum dynamical activity [\Eref{Nishiyama_homodyne_QDA}].

\section{Derivation of Quantum Dynamical Activity under Feedback Control by Gaussian Measurement}\label{gaussian_derivation}
In this section, we provide the derivation of quantum dynamical activity under feedback control by gaussian measurement in two forms[\Eref{Nakajima_gaussian_QDA}, \Eref{Nishiyama_gaussian_QDA}] based on the methods in Ref. \cite{nakajimaSymmetriclogarithmicderivativeFisherInformation2023, nishiyamaExactSolutionQuantum2024}. 

At first, we derive NU-type of quantum dynamical activity [\Eref{Nakajima_gaussian_QDA}]. When we define $C(\theta, \phi)$ as
\begin{equation}\label{def_gaussian_c}
    C(\theta, \phi)\equiv \mathrm{Tr}_S[\rho^{\theta, \phi}(\tau)],
\end{equation}
where $\rho^{\theta, \phi}(\tau)$ follows \Eref{gaussian_two_sided}, quantum dynamical activity $\mathcal{B}_{\mathrm{gau}}^{\mathrm{fb}}(\tau)$ we want can be described as follows:
\begin{equation}\label{what_calc_gaussian}
    \mathcal{B}_{\mathrm{gau}}^{\mathrm{fb}}(\tau)=4[\partial_\theta \partial_\phi C(\theta, \phi)-\partial_\theta C(\theta, \phi) \partial_\phi C(\theta, \phi)]|_{\theta=\phi=0}.
\end{equation}
From \Eref{liouville_evolve} and \Eref{def_gaussian_c}, we obtain
\begin{equation}
    C(\theta, \phi)=\langle\!\langle1|\exp({\hat{\mathcal{L}}_\mathrm{G}^{\theta, \phi}}\tau)|\rho^{\theta, \phi}(0)\rangle\!\rangle.
\end{equation}
The first derivative of $C(\theta, \phi)$ becomes
\begin{equation}\label{first_der_g}
    \partial_{\theta_i}C(\theta, \phi)=\int_0^\tau du\langle\!\langle1|\exp(\hat{\mathcal{L}}_\mathrm{G}^{\theta, \phi}(\tau-u))\partial_{\theta_i}\hat{\mathcal{L}}_\mathrm{G}^{\theta, \phi}\exp(\hat{\mathcal{L}}_\mathrm{G}^{\theta, \phi}u)|\rho^{\theta, \phi}(0)\rangle\!\rangle,
\end{equation}
where $\theta_i=\theta, \phi$. From \Eref{liouville_conserve2}, $\partial_{\theta_i}C(\theta, \phi)|_{\theta=\phi=0}$ can be written by
\begin{equation}
    \partial_{\theta_i}C(\theta, \phi)|_{\theta=\phi=0}=\int_0^\tau ds \langle\!\langle1|\partial_{\theta_i} \hat{\mathcal{L}}_\mathrm{G}^{\theta, \phi}|\rho(s)\rangle\!\rangle|_{\theta=\phi=0}.
\end{equation}
By calculating this, we can obtain
\begin{equation}
    \partial_{\theta}C(\theta, \phi)|_{\theta=\phi=0}=\int_0^\tau ds \mathrm{Tr}_S[\mathcal{K}_{\mathrm{G}1}\rho(s)],
\end{equation}
\begin{equation}
    \partial_{\phi}C(\theta, \phi)|_{\theta=\phi=0}=\int_0^\tau ds \mathrm{Tr}_S[\mathcal{K}_{\mathrm{G}2}\rho(s)],
\end{equation}
where $\mathcal{K}_{\mathrm{G}1}$ and $\mathcal{K}_{\mathrm{G}2}$ are defined in \Eref{k_g1} and \Eref{k_g2}. Then we can calculate the second term of \Eref{what_calc_gaussian} as follows:
\begin{equation}\label{gaussian_qda_2}
\begin{split}
    &-\partial_\theta C(\theta, \phi)\partial_\phi C(\theta, \phi)|_{\theta=\phi=0}\\
    &=-\Pi_{i=1}^2 \int_0^\tau ds \mathrm{Tr}[\mathcal{K}_{\mathrm{G}i}\rho(s)]\\
    &=\left\{-i\int_0^\tau ds \big(\mathrm{Tr}_S[H\rho(s)]+\frac{1}{4}\mathrm{Tr}_S[F\rho(s)Y+FY\rho(s)]\big)\right\}\times\left\{i\int_0^\tau ds \big(\mathrm{Tr}_S[H\rho(s)]+\frac{1}{4}\mathrm{Tr}_S[F\rho(s)Y+FY\rho(s)]\big)\right\}\\
    &= -\left\{\int_0^\tau ds \big(\mathrm{Tr}_S[H\rho(s)]+\frac{1}{4}\mathrm{Tr}_S[F\rho(s)Y+FY\rho(s)]\big)\right\}^2.
\end{split}
\end{equation}
From \Eref{first_der_g}, the first term of \Eref{what_calc_gaussian} becomes
\begin{equation}
\begin{split}\label{second_der_g}
    \partial_{\theta}\partial_{\phi}C(\theta, \phi)|_{\theta=\phi=0}=&\int_0^\tau du \langle\!\langle1|\partial_\theta\exp(\hat{\mathcal{L}}_\mathrm{G}^{\theta, \phi}(\tau-u))\partial_{\phi}\hat{\mathcal{L}}_\mathrm{G}^{\theta, \phi}\exp(\hat{\mathcal{L}}_\mathrm{G}^{\theta, \phi}u)\\
    &+\exp(\hat{\mathcal{L}}_\mathrm{G}^{\theta, \phi}(\tau-u))\partial_\theta\partial_{\phi}\hat{\mathcal{L}}_\mathrm{G}^{\theta, \phi}\exp(\hat{\mathcal{L}}_\mathrm{G}^{\theta, \phi}u)\\
    &+\exp(\hat{\mathcal{L}}_\mathrm{G}^{\theta, \phi}(\tau-u))\partial_{\phi}\hat{\mathcal{L}}_\mathrm{G}^{\theta, \phi}\partial_\theta\exp(\hat{\mathcal{L}}_\mathrm{G}^{\theta, \phi}u)|\rho^{\theta, \phi}(0)\rangle\!\rangle|_{\theta=\phi=0}.
\end{split}
\end{equation}
The second term of \Eref{second_der_g} can be calculated as follows
\begin{equation}\label{second_term_g}
    \int_0^\tau du\langle\!\langle1|\partial_\theta\partial_{\phi}\hat{\mathcal{L}}_\mathrm{G}^{\theta, \phi}|\rho^{\theta, \phi}(s)\rangle\!\rangle|_{\theta=\phi=0}=\frac{1}{4}\mathcal{A}_{\mathrm{gau}}^{\mathrm{fb}}(\tau).
\end{equation}
The first term of \Eref{second_der_g} becomes
\begin{equation}\label{first_term_g}
    \int_0^\tau du \langle\!\langle1|\int_u^\tau ds \exp(\hat{\mathcal{L}}_\mathrm{G}^{\theta, \phi}(\tau-s))\partial_{\theta}\hat{\mathcal{L}}_\mathrm{G}^{\theta, \phi}
    \exp(\hat{\mathcal{L}}_\mathrm{G}^{\theta, \phi}(s-u))\partial_{\phi}\hat{\mathcal{L}}_\mathrm{G}^{\theta, \phi}\exp(\hat{\mathcal{L}}_\mathrm{G}^{\theta, \phi}(u))|\rho^{\theta, \phi}(0)\rangle\!\rangle|_{\theta=\phi=0}
    =I_{\mathrm{G}2}.
\end{equation}
The third term of \Eref{second_der_g} becomes
\begin{equation}\label{third_term_g}
    \int_0^\tau du \langle\!\langle1|\exp(\hat{\mathcal{L}}_\mathrm{G}^{\theta, \phi}(\tau-u))\partial_{\phi}\hat{\mathcal{L}}_\mathrm{G}^{\theta, \phi}
    \int_0^u ds \exp(\hat{\mathcal{L}}_\mathrm{G}^{\theta, \phi}(s-u))\partial_{\theta}\hat{\mathcal{L}}_\mathrm{G}^{\theta, \phi}\exp(\hat{\mathcal{L}}_\mathrm{G}^{\theta, \phi}(u))|\rho^{\theta, \phi}(0)\rangle\!\rangle|_{\theta=\phi=0}
    =I_{\mathrm{G}1}.
\end{equation}
Then, we obtain the first term of \Eref{what_calc_gaussian} as follows:
\begin{equation}\label{gaussian_qda_1}
    \partial_{\theta}\partial_{\phi}C(\theta, \phi)|_{\theta=\phi=0}=\frac{1}{4}\mathcal{A}_{\mathrm{gau}}^{\mathrm{fb}}(\tau)+I_{\mathrm{G}1}+I_{\mathrm{G}2}.
\end{equation}
From \Eref{gaussian_qda_1} and \Eref{gaussian_qda_2}, we obtain NU-type of quantum dynamical activity [\Eref{Nakajima_gaussian_QDA}].

Next, we can derive NH-type of quantum dynamical activity [\Eref{Nishiyama_gaussian_QDA}] from NU-type of quantum dynamical activity. From the cyclic property of trace, we can obtain the following relations:
\begin{equation}
    \mathrm{Tr}_S[\mathcal{K}_{\mathrm{G}1}\bullet]=-i(\mathrm{Tr}_S[H\bullet]+\frac{1}{4}\mathrm{Tr}_S[F\bullet Y+FY\bullet]),
\end{equation}
\begin{equation}
    \mathrm{Tr}_S[\mathcal{K}_{\mathrm{G}1}\bullet]=i(\mathrm{Tr}_S[H\bullet]+\frac{1}{4}\mathrm{Tr}_S[F\bullet Y+FY\bullet]).
\end{equation}
By applying these relations, $I_{\mathrm{G}1}$ and $I_{\mathrm{G}2}$ become
\begin{equation}
\begin{split}
    I_{\mathrm{G}1}=&i\int_0^\tau ds \int_0^s du \Big\{\mathrm{Tr}_S[H\exp(\mathcal{L}_\mathrm{G}(s-u))\mathcal{K}_{\mathrm{G}1}\rho(u)]\\
    &+\frac{1}{4}\mathrm{Tr}[F\exp(\mathcal{L}_{\mathrm{G}}(s-u))(\mathcal{K}_{\mathrm{G}1}\rho(u))Y+FY\exp(\mathcal{L}_{\mathrm{G}}(s-u))(\mathcal{K}_{\mathrm{G}1}\rho(u))]\Big\}
\end{split}
\end{equation}
\begin{equation}
\begin{split}
    I_{\mathrm{G}2}=&-i\int_0^\tau ds \int_0^s du \Big\{\mathrm{Tr}_S[H\exp(\mathcal{L}_\mathrm{G}(s-u))\mathcal{K}_{\mathrm{G}2}\rho(u)]\\
    &+\frac{1}{4}\mathrm{Tr}[F\exp(\mathcal{L}_{\mathrm{G}}(s-u))(\mathcal{K}_{\mathrm{G}2}\rho(u))Y+FY\exp(\mathcal{L}_{\mathrm{G}}(s-u))(\mathcal{K}_{\mathrm{G}2}\rho(u))]\Big\}.
\end{split}
\end{equation}
By representing $\exp(\mathcal{L}_\mathrm{G}(s-u))$ with Kraus operators $M_z$ and $U_z$ and using cyclic property of trace, we have
\begin{equation}
\begin{split}
    I_{\mathrm{G}1}&=i\int_0^\tau ds \int_0^{s} du \left\{\operatorname{Tr}_S[\overline{H}(s - u)\mathcal{K}_{\mathrm{G1}} \rho(u)]+ \frac{1}{4}\mathrm{Tr}_S[\overline{YF}(s-u)(\mathcal{K}_{\mathrm{G}1}\rho(u))]+ \frac{1}{4}\mathrm{Tr}_S[\overline{FY}(s-u)(\mathcal{K}_{\mathrm{G}1}\rho(u))] \right\},
\end{split}
\end{equation}
\begin{equation}
\begin{split}
    I_{\mathrm{G}2}&=-i\int_0^\tau ds \int_0^{s} du \left\{\operatorname{Tr}_S[\overline{H}(s - u)\mathcal{K}_{\mathrm{G2}} \rho(u)]+ \frac{1}{4}\mathrm{Tr}_S[\overline{YF}(s-u)(\mathcal{K}_{\mathrm{G}2}\rho(u))]+ \frac{1}{4}\mathrm{Tr}_S[\overline{FY}(s-u)(\mathcal{K}_{\mathrm{G}2}\rho(u))] \right\},
\end{split}
\end{equation}
where $\overline{\bullet}$ is given by
\begin{equation}
    \overline{\bullet}=\sum_{\bm{z}} M_{z_0}^\dagger U_{z_0}^\dagger\cdots M_{z_{N_\tau-1}}^\dagger U_{z_{N_\tau-1}}^\dagger\bullet U_{z_{N_\tau-1}}M_{z_{N_\tau-1}}\cdots U_{z_0}M_{z_0},
\end{equation}
when time interval $[u, s]$ is divided by large $N_\tau$. Given that $\bullet(t)$ evolves by Kraus operator $U_z^\dagger$ and $M_z^\dagger$ during infinitesimal time to obtain the concrete form of $\overline{\bullet}$, we obtain
\begin{equation}
\begin{split}
    \bullet(t+dt)&=\int dz M_z^\dagger e^{I_\mathrm{gau}(t)\mathcal{F}^\dagger dt}e^{\mathcal{H}^\dagger dt}\bullet(t) M_z\\
    &=\int dz M_z^\dagger e^{\langle Y \rangle \mathcal{F}^\dagger dt} e^{\frac{\Delta W}{2\sqrt{\lambda}} \mathcal{F}^\dagger}e^{\mathcal{H}^\dagger dt}\bullet(t) M_z\\
    &=\int dz M_z^\dagger \left(1 + \langle Y \rangle \mathcal{F}^\dagger dt \right)\left(1 + \frac{\Delta W}{2\sqrt{\lambda}} \mathcal{F}^\dagger + \frac{dt}{8\lambda} \left(\mathcal{F}^\dagger\right)^2\right)\left(1 + \mathcal{H}^\dagger dt \right)\bullet(t) M_z+o(dt)\\
    &=\int dz M_z^\dagger \left(1+\frac{\Delta W}{2\sqrt{\lambda}}\mathcal{F}
    ^\dagger+\frac{dt}{8\lambda}(\mathcal{F}^\dagger)^2+\langle Y\rangle \mathcal{F}^\dagger dt\right)\left(1 + \mathcal{H}^\dagger dt \right)\bullet(t) M_z+o(dt)\\
    &=\int dz  M_z^\dagger \left(1+\mathcal{H}^\dagger dt+\langle Y \rangle\mathcal{F}^\dagger dt+\frac{\Delta W}{2\sqrt{\lambda}}\mathcal{F}^\dagger+\frac{dt}{8\lambda}(\mathcal{F}^\dagger)^2\right)\bullet(t) M_z+o(dt),
\end{split}
\end{equation}
where $dW$ is replaced by $\Delta W$ for clarity of the equation. When we convert the integral by $I_\mathrm{gau}(t)$ to $\Delta W$, the following relation holds:
\begin{equation}
    dI_\mathrm{gau}(t) = \frac{d \Delta W}{2\sqrt{\lambda}dt}.
\end{equation}
Thus, we can calculate as follows:
\begin{equation}
\begin{split}
    \bullet(t+dt)&=\frac{1}{2\sqrt{\lambda}dt}\left(\frac{2\lambda dt}{\pi}\right)^{\frac{1}{2}}\\
    &\times \left[\sum_{y, y'}\int d\Delta We^{-\lambda dt(\langle Y\rangle+\frac{\Delta W}{2\sqrt{\lambda}dt}-y)^2}e^{-\lambda dt(\langle Y\rangle+\frac{\Delta W}{2\sqrt{\lambda}dt}-y')^2}\langle y|(1+\mathcal{H}^\dagger dt+\langle Y \rangle\mathcal{F}^\dagger dt+\frac{dt}{8\lambda}(\mathcal{F}^\dagger)^2)\bullet(t)|y'\rangle|y\rangle\langle y'|\right]\\
    &\times \left[\sum_{y, y'}\int d\Delta We^{-\lambda dt(\langle Y\rangle+\frac{\Delta W}{2\sqrt{\lambda}dt}-y)^2}e^{-\lambda dt(\langle Y\rangle+\frac{\Delta W}{2\sqrt{\lambda}dt}-y')^2}\langle y|\frac{\Delta W}{2\sqrt{\lambda}}\mathcal{F}^\dagger\bullet(t)|y'\rangle|y\rangle\langle y'|\right]+o(dt)\\
    &=\mathcal{G}(\bullet+\mathcal{H}^\dagger \bullet dt+\langle Y\rangle \mathcal{F}^\dagger\bullet dt+\frac{(\mathcal{F}^\dagger)^2}{8\lambda}\bullet dt)+\sum_{y, y'}\frac{dt}{2}(y'+y-2\langle Y\rangle)\langle y|\mathcal{F}^\dagger\bullet(t)|y'\rangle|y\rangle\langle y'|+o(dt),
\end{split}
\end{equation}
where $\mathcal{G}$ is defined by
\begin{equation}\label{G_def}
    \mathcal{G}\bullet\equiv 1-\frac{\lambda}{2}\bullet Y^2 dt-\frac{\lambda}{2}Y^2\bullet dt+\lambda Y\bullet Y dt.
\end{equation}
\EEref{G_def} is obtained by the following relations:
\begin{equation}
\begin{split}
    \bullet Y^2 &= \sum_{y, y'} \langle y | \bullet | y' \rangle | y \rangle \langle y' | \sum_{y''} y''^2 | y'' \rangle \langle y'' | \\
             &= \sum_{y, y'} y'^{2} \langle y | \bullet | y' \rangle | y \rangle \langle y' |, \\
    Y^2 \bullet &= \sum_{y''} y''^2 | y'' \rangle \langle y'' | \sum_{y, y'} \langle y | \bullet | y' \rangle | y \rangle \langle y' | \\
             &= \sum_{y, y'} y^2 \langle y | \bullet | y' \rangle | y \rangle \langle y' |, \\
    Y \bullet Y &= \sum_{y''} y'' | y'' \rangle \langle y'' | \sum_{y, y'} \langle y | \bullet | y' \rangle | y \rangle \langle y' | \sum_{y'''} y''' | y''' \rangle \langle y''' | \\
             &= \sum_{y, y'} y y' \langle y | \bullet | y' \rangle | y \rangle \langle y' |.
\end{split}
\end{equation}
Further calculations show that
\begin{equation}
\begin{split}
\bullet(t+dt)=&\bullet(t)+\mathcal{H}^\dagger \bullet(t)dt+\langle Y\rangle\mathcal{F}^\dagger\bullet(t)dt+\frac{(\mathcal{F}^\dagger)^2}{8\lambda}\bullet(t) dt-\frac{\lambda}{2}\bullet(t) Y^2dt-\frac{\lambda}{2}Y^2\bullet(t) dt+\lambda Y\bullet(t) Y dt\\
&+\sum_{y, y'}\frac{dt}{2}(y'+y)\langle y|\mathcal{F}^\dagger \bullet(t)|y'\rangle|y\rangle\langle y'|-\langle Y\rangle\mathcal{F}^\dagger \bullet(t) dt+o(dt).
\end{split}
\end{equation}
By using the following relations:
\begin{equation}
\begin{split}
        \bullet Y &= \sum_{y, y'} \langle y | \bullet | y' \rangle | y \rangle \langle y' | \sum_{y''} y'' | y'' \rangle \langle y'' | \\
             &= \sum_{y, y'} y' \langle y | \bullet | y' \rangle | y \rangle \langle y' |, \\
    Y \bullet &= \sum_{y''} y'' | y'' \rangle \langle y'' | \sum_{y, y'} \langle y | \bullet | y' \rangle | y \rangle \langle y' | \\
             &= \sum_{y, y'} y \langle y | \bullet | y' \rangle | y \rangle \langle y' |, \\
\end{split}
\end{equation}
we can obtain
\begin{equation}
    \bullet(t+dt)=\bullet(t)+\mathcal{H}^\dagger\bullet(t) dt+\lambda Y\bullet(t) Y dt-\frac{\lambda}{2}\bullet(t) Y^2 dt -\frac{\lambda}{2}Y^2\bullet(t) dt+\frac{1}{2}(\mathcal{F}^\dagger(\bullet) Y+Y\mathcal{F}^\dagger (\bullet))dt+\frac{(\mathcal{F}^\dagger)^2}{8\lambda}\bullet(t) dt+o(dt).
\end{equation}
Then we can derive the following equation:
\begin{equation}\label{gaussian_dag_evolve}
    \dot{\bullet}= i[H, \bullet]+\lambda Y\bullet Y-\frac{\lambda}{2}\{Y^2, \bullet\}+\frac{1}{2}(\mathcal{F}^\dagger (\bullet) Y+Y\mathcal{F}^\dagger (\bullet))+\frac{(\mathcal{F}^\dagger)^2}{8\lambda}\bullet \equiv\mathcal{L}_{\mathrm{G}}^\dagger\bullet.
\end{equation}
From \Eref{gaussian_dag_evolve}, we can get the definition of $\overline{\bullet}$ as follows:
\begin{equation}
    \overline{\bullet}(t)\equiv \exp(\mathcal{L}_{\mathrm{G}}^\dagger t)\bullet.
\end{equation}
By using the following relation:
\begin{equation}
    i\mathcal{K}_{\mathrm{G}1}\bullet-i\mathcal{K}_{\mathrm{G}2}\bullet=H\bullet+\bullet H-i\frac{\lambda}{2}Y^2\bullet+i\frac{\lambda}{2}\bullet Y^2+\frac{1}{2}FY\bullet+\frac{1}{2}\bullet YF-\frac{i}{8\lambda}F^2\bullet+\frac{i}{8\lambda}\bullet F,
\end{equation}
and $(\overline{YF})^\dagger=\overline{FY}$, we can obtain $I_{\mathrm{G}1} + I_{\mathrm{G}2}$ as follows:
\begin{equation}
    I_{\mathrm{G}1} + I_{\mathrm{G}2} = \int_0^\tau ds \int_0^{s} du 2\mathrm{Re}\left(\operatorname{Tr}_S\left[\overline{H}(s - u)\rho(u)H_{\text{eff}}^{\mathrm{g}\dagger}+\frac{1}{2}\mathrm{Tr}_S[\{\overline{YF}(s-u)+\overline{FY}(s-u)\}\rho(u)H_{\text{eff}}^{\mathrm{g}\dagger}]  \right]\right).
\end{equation}
Then, we can get NH-type of quantum dynamical activity [\Eref{Nishiyama_gaussian_QDA}].

\section{Taylor Expansion of Quantum Dynamical Activity under Feedback Control by Jump Measurement}\label{jump_taylor_derivation}
In this section, to analytically clarify the contribution of feedback control to the quantum dynamical activity under jump measurement $\mathcal{B}_\mathrm{jmp}^\mathrm{fb}(\tau)$, we perform a Taylor expansion of $\mathcal{B}_\mathrm{jmp}^\mathrm{fb}(\tau)$ and derive \Eref{jump_taylor}. This analysis is conducted under the assumption of a short time interval ($\tau \ll 1$) and a weak feedback regime ($\nu_z\ll 1$). Consequently, all terms of second order and higher in $\tau$ and $\nu_z$ are neglected in the following derivation. $\mathcal{B}_\mathrm{jmp}^\mathrm{fb}(\tau)$ is represented as follows [\Eref{Nishiyama_jump_QDA}] :

\begin{equation}
    \mathcal{B}_{\mathrm{jmp}}^{\mathrm{fb}}(\tau) = \mathcal{A}(\tau) + 8 \int_0^\tau ds_1 \int_0^{s_1} ds_2 \, \mathrm{Re}\Big(\mathrm{Tr}_S[H_\mathrm{eff}^\dagger \breve{H}(s_1 - s_2) \rho(s_2) ] \Big) - 4 \left( \int_0^\tau ds \, \mathrm{Tr}[H \rho(s)] \right)^2,
\end{equation}
where
\begin{equation}
\begin{split}
    &\breve{H}(t)\equiv e^{\mathcal{L}_{\mathrm{J}}^\dagger t}H, \\
    &\dot{\mathcal{O}}=\mathcal{L}_{\mathrm{J}}^\dag\mathcal{O}\equiv i[H, \mathcal{O}]+\sum_{z=1}^{N_c}L_z^\dag e^{\nu_z \mathcal{F}^\dagger}(\mathcal{O})L_z-\frac{1}{2}\{L_z^\dag L_z, \mathcal{O}\}, \\
    & e^{\nu_z \mathcal{F}^\dagger}(\mathcal{O})=e^{i\nu_z F}\mathcal{O}e^{-i\nu_zF}.
\end{split}
\end{equation}
Under the conditions $\tau\ll 1$ and $\nu_z\ll 1$, the following expression is obtained:
\begin{equation}
\begin{split}
    &\breve{H}(t)\simeq(1+\mathcal{L}_\mathrm{J}t)H,\\
    &e^{i\nu_z F}\mathcal{O}e^{-i\nu_z F}\simeq\mathcal{O}+i\nu_z[F, \mathcal{O}].
\end{split}
\end{equation}
Thus, we obtain
\begin{equation}
    \breve{H}(t)\simeq 1+\mathcal{L}^\dagger Ht+\sum_z i\nu_zL_z[F, H]L_z^\dagger t,
\end{equation}
where $\mathcal{L}$ is the adjoint superoperator for the dynamics without feedback control, i.e., Lindblad dynamics. Similarly, in the short-time limit ($\tau\ll 1$), the quantum dynamical activity in the absence of feedback control $\mathcal{B}(\tau)$ is given by
\begin{equation}
    \mathcal{B}(\tau) = \mathcal{A}(\tau) + 8 \int_0^\tau ds_1 \int_0^{s_1} ds_2 \, \mathrm{Re}\Big(\mathrm{Tr}_S[H_\mathrm{eff}^\dagger \big(1+\mathcal{L}^\dagger H(s_1-s_2)\big) \rho(s_2) ] \Big) - 4 \left( \int_0^\tau ds \, \mathrm{Tr}_S[H \rho(s)] \right)^2.
\end{equation}
Therefore, $\mathcal{B}_{\mathrm{jmp}}^{\mathrm{fb}}(\tau)$ can be written by
\begin{equation}
    \mathcal{B}_{\mathrm{jmp}}^{\mathrm{fb}}(\tau) \simeq \mathcal{B}(\tau)+8 \int_0^\tau ds_1\int_0^{s_1} ds_2 \mathrm{Re}\Big\{\mathrm{Tr}_S\big[H_\mathrm{eff}^\dagger \Big(\sum_z i\nu_zL_z^\dagger[F, H] L_z\Big)(s_1 - s_2) \rho(s_2) \big] \Big\}.
\end{equation}
Then, we can express $\mathcal{B}_{\mathrm{jmp}}^{\mathrm{fb}}(\tau)$, as the sum of $\mathcal{B}(\tau)$ and a term representing feedback contribution, as shown in \Eref{jump_taylor}.

\end{widetext}

%

\end{document}